\documentclass[twocolumn,pra,aps,superscriptaddress,showpacs,eqsecnum]{revtex4}

\usepackage{yfonts} 
\usepackage{amsmath,amsfonts,mathrsfs,amssymb}
\usepackage{tikz}
\usepackage{graphicx}
\usepackage{epstopdf}
\usepackage{epsfig}
\usepackage{color}
\usepackage{wasysym}
\usepackage[breaklinks=true]{hyperref}
\usepackage{url}

\newcommand{\bra}[1]{\left\langle{#1}\right|}
\newcommand{\ket}[1]{\left|{#1}\right\rangle}
\newcommand{\op}[2]{\ket{#1}\!\bra{#2}}
\newcommand{\expt}[1]{\left\langle{#1}\right\rangle}
\newcommand{\ip}[2]{\left\langle{#1}\right|\left.{#2}\right\rangle}

\newcommand{\sE}{\mathcal{E}}

\newcommand{\sA}{\mathcal{A}}
\newcommand{\sR}{\mathcal{R}}

\newcommand{\mfA}{\mbox{\textgoth{A}}}

\newcommand{\tr}{\mbox{tr}}

\newcommand{\SNRone}{\mbox{SNR}_1}
\newcommand{\SNRtwo}{\mbox{SNR}_2}
\newcommand{\SNRnum}{\mbox{SNR}_{N}}

\newcommand{\mytilde}{\raise.17ex\hbox{$\scriptstyle\mathtt{\sim}$}}
\newcommand{\erf}[1]{Eq.~(\ref{#1})}
\newcommand{\frf}[1]{Fig.~\ref{#1}}
\newcommand{\srf}[1]{Sec.~\ref{#1}}
\newcommand{\dg}{^\dag}

\def\Echeck{\mathcal{E}_{\checkmark}}
\def\Pcheck{p_{\checkmark}}

\begin{document}
\pacs{42.65.Yj, 03.67.-a, 42.50.Lc}
\title{Quantum limits on probabilistic amplifiers}

\author{Shashank Pandey}
\affiliation{Center for Quantum Information and Control, University
of New Mexico, Albuquerque, NM 87131-0001, USA}
\author{Zhang Jiang}
\affiliation{Center for Quantum Information and Control, University
of New Mexico, Albuquerque, NM 87131-0001, USA}
\author{Joshua Combes}
\affiliation{Center for Quantum Information and Control, University
of New Mexico, Albuquerque, NM 87131-0001, USA}
\author{Carlton M. Caves}
\email{ccaves@unm.edu}
\affiliation{Center for Quantum Information
and Control, University of New Mexico, Albuquerque, NM 87131-0001,
USA}
\affiliation{Centre for Engineered Quantum Systems, School of
Mathematics and Physics, University of Queensland, Brisbane,
Queensland 4072, Australia}

\begin{abstract}
An ideal phase-preserving linear amplifier is a deterministic device
that adds to an input signal the minimal amount of noise consistent
with the constraints imposed by quantum mechanics. A noiseless linear
amplifier takes an input coherent state to an amplified coherent
state, but only works part of the time. Such a device is actually
better than noiseless, since the output has less noise than the
amplified noise of the input coherent state; for this reason we refer
to such devices as {\em immaculate}.  Here we bound the working
probabilities of probabilistic and approximate immaculate amplifiers
and construct theoretical models that achieve some of these bounds.
Our chief conclusions are the following: (i)~the working probability
of any phase-insensitive immaculate amplifier is very small in the
phase-plane region where the device works with high fidelity;
(ii)~phase-sensitive immaculate amplifiers that work only on coherent
states sparsely distributed on a phase-plane circle centered at the
origin can have a reasonably high working probability.
\end{abstract}

\date{\today}

\maketitle

\section{Introduction and motivation}
\label{sec:intro}

Classical non-inverting amplifiers take a macroscopic input signal,
such as a time-varying voltage, and produce an output signal that is
a rescaled version of the input signal.  The ratio of the input
amplitude to the output amplitude is called the gain $g$ of the
amplifier.  Classical amplifiers are used ubiquitously, e.g., to
boost signal strength for classical communications or to increase the
power of signals driving loud speakers.  In principle, a classical
amplifier can be noise free in the sense that no noise is added to
the input signal. The only truly fundamental limit on amplification
comes from quantum mechanics.

The canonical quantum amplifier is called a phase-preserving linear
quantum amplifier.  It takes an input bosonic signal and produces a
larger output signal~\cite{Haus1962a,Caves1982a,CavComJiaPan12},
while preserving the phase.  The quantum constraints on the operation
of such a device are ultimately a consequence of unitarity and can be
thought as coming from the prohibition on transformations that
increase the distinguishability of nonorthogonal
states~\cite{WooZur82,Dieks82}.  The quantum constraint on a
high-gain device can be expressed as the requirement that the
amplifier must add noise that, when referred to the input, is at
least as big as an extra unit of vacuum noise.  A device that
achieves the minimal added noise is called an {\em ideal\/} linear
amplifier.

To understand the purpose of quantum amplifiers, it is instructive to
look at how they are used.  An illustrative case involves experiments
probing quantum mechanics at microwave frequencies.  Experimenters
wish to measure the small amplitude and phase shifts of a field that
is used to probe another quantum system.  It turns out that
quantum-limited simultaneous measurements of both amplitude and phase
shifts introduce the same additional unit of vacuum noise as does an
ideal linear amplifier~\cite{Arthurs1965a}.  Thus, in principle,
measuring at the input or amplifying and measuring at the output both
provide the same signal-to-noise ratio (SNR); the practical question
becomes whether it is easier to do quantum-limited measurement or to
do quantum-limited amplification and subsequent measurement at the
output.  The answer at microwave frequencies is that amplifiers
operate closer to quantum limits.

Recently Ralph and Lund~\cite{RalLu09} proposed a device, which they
call a ``nondeterministic noiseless linear amplifier,'' previously considered 
by Fiur\'a\v{s}ek~\cite{Fiur04a} in the context of probabilistic cloning. The idea
behind the Ralph-Lund device is that it might be possible to {\em
improve\/} the SNR in some number of trials/experiments, while the
device fails in the remaining runs. Specifically, what Ralph and Lund
proposed is a device that takes an input coherent state
$|\alpha\rangle$ to a target coherent state $|g\alpha\rangle$ with
(success) probability $\Pcheck$ and fails with probability
$1-\Pcheck$. Such a device is even better than noiseless, because
when the output noise is referred to the input, it is smaller than
the original coherent-state noise by a factor of $1/g^2$.  In
particular, it is better than a device that amplifies the input noise
to the output without the addition of any noise, a device that we
call a {\em perfect amplifier}.  Because it is better than perfect,
we call Ralph and Lund's proposal an {\em immaculate amplifier}.  The
purpose of this paper is to analyze in detail and to bound the
performance of immaculate linear amplifiers.

In Sec.~\ref{sec:laprior} we review recent work on deterministic
linear amplifiers~\cite{CavComJiaPan12}, which allows us to consider
on the same footing ideal linear amplifiers and (unphysical) perfect
and immaculate amplifiers. We use this discussion to motivate the
idea of nondeterministic, or probabilistic, versions of perfect and
immaculate amplifiers, and we use a simple uncertainty-principle
argument to bound the working probability of probabilistic perfect
and immaculate amplifiers.

Section~\ref{sec:relation_cloning} reviews the relation between
amplification and cloning, thus connecting the results in this paper
to the literature on cloning of coherent states, and
Sec.~\ref{sec:paprior} reviews proposals for and experimental
implementations of immaculate linear amplifiers.

Sections~\ref{sec:usd} and~\ref{sec:kraus} are the heart of the
paper, the place where we derive bounds on the operation of
immaculate amplifiers. Immaculate amplifiers that produce
the target coherent state exactly, but are allowed to fail, are the
subject of Sec.~\ref{sec:usd}; they are closely related to
unambiguous state discrimination~\cite{Che98,CheBar98}, in which one
discriminates among a set of linearly independent states exactly, but
can declare a failure to discriminate.  We use results from
unambiguous state discrimination to bound the working probability of
an immaculate amplifier that amplifies $M$ coherent states uniformly
spaced around a circle of radius $|\alpha|$ centered at the origin of
the phase plane. In the case of many coherent states on both the
input and output circles, i.e, assuming $M\gg g^2|\alpha|^2$, the
working probability is bounded by
\begin{align}
\Pcheck\le\frac{e^{(g^2-1)|\alpha|^2}}{g^{2(M-1)}}
\ll\left(\frac{\sqrt e}{g^2}\right)^{2(M-1)}\,.
\label{eq:pmanymany}
\end{align}
This success probability decreases exponentially with $M$
and goes to zero in the phase-insensitive limit $M\rightarrow\infty$.
We stress that this means that an immaculate amplifier that works
exactly on an entire circle of input coherent states never works.

For an immaculate amplifier that acts on all coherent states on $M$
equally spaced spokes of a disk of any radius {$|\alpha|>0$}
centered at the origin, the success probability is governed by the
limiting circle of zero radius and thus is bounded by
\begin{align}
\Pcheck\le\frac{1}{g^{2(M-1)}}
\label{eq:pdisk}
\end{align}
for any $M\ge2$. This success probability goes to zero in
the phase-insensitive limit $M\rightarrow\infty$.

On a more optimistic note, we also show in Sec.~\ref{sec:usd} that if
the $M$ coherent states are more than about a vacuum unit apart on
the input circle, they can be immaculately amplified with a success
probability exceeding a half.  This suggests that practical
applications of immaculate amplifiers are likely to be as amplifiers
that are both phase sensitive and amplitude specific in that they
only work well on a discrete set of states on a particular
phase-plane circle. Such an amplitude-specific, phase-sensitive
amplifier might prove useful, for example, in discriminating the
coherent states used in phase-shift keying~\cite{NaiYenGuh12,Bec13}.

The results of Sec.~\ref{sec:usd} indicate that exact immaculate
amplification and phase insensitivity don't go well together.  In
Sec.~\ref{sec:kraus} we explore this incompatibility further by
dropping exactness and investigating the performance of approximate,
probabilistic immaculate amplifiers that are explicitly phase
insensitive.  We characterize such a device by its amplitude gain and
by the radius $\sqrt N/g$ of the disk, centered at the origin, over
which it amplifies an input coherent $|\alpha\rangle$ to the target
output state $|g\alpha\rangle$ with near unit fidelity. The
high-fidelity outputs thus lie within a disk of radius $\sqrt N$. By
finding the optimal such amplifier, we show that the best success
probability in the high-fidelity input region is
\begin{equation}
\Pcheck=\frac{e^{-|\alpha|^2}}{g^{2N}}\,,
\quad
|\alpha|^2\alt N/g^2\,,
\label{eq:pintro}
\end{equation}
which decreases exponentially with $N$.  We use our results to
investigate the performance of phase-insensitive immaculate
amplifiers within the context of the signal-to-noise ratios for
measurements of amplitude and phase shifts discussed above.

Because the success probability~\ref{eq:pintro}) is so small, we
suggest that a good performance measure for phase-insensitive
immaculate amplifiers must include both the fidelity with the target
output $|g\alpha\rangle$ and the success probability.  A natural
combination is the product of the two, which can be thought of as the
overall probability to reach the target.  We show that over the whole
range of operation of the optimal phase-insensitive immaculate
amplifier, this probability-fidelity product is never better than
that of the identity operation.  This can be summarized by saying
that in terms of the probability-fidelity product, {\em
phase-preserving\/} immaculate amplification is never better than
doing nothing, thus re{\"e}nforcing our conclusion that any practical
application of immaculate amplification lies in phase-sensitive
amplification.

{A concluding Sec.~\ref{sec:con} wraps up by summarizing our
key results and discussing avenues along which future research might
and should proceed.}

\section{Physical and unphysical linear amplifiers}
\label{sec:laprior}

\subsection{Context}
\label{subsec:context}

The setting for our investigation is a signal carried by a
single-mode field,
\begin{align}
E(t)&\!=\!\frac{1}{2}(ae^{-i\omega t}+a^\dagger e^{-i\omega t})
\!=\!\frac{1}{\sqrt2}(x_1\cos\omega t+x_2\sin\omega t)\,.
\end{align}
This {\em primary mode}, which we label by $A$, is to undergo
phase-preserving linear amplification.  The annihilation and creation
operators, $a$ and $a^\dagger$, are related to the Hermitian
quadrature components, $x_1$ and $x_2$, by $a=(x_1+ix_2)/\sqrt{2}$,
$a^\dagger =(x_1-ix_2)/\sqrt{2}$, where $[a,a^\dagger]=1$ or,
equivalently, $[x_1,x_2]=i$.

The annihilation operator is a complex-amplitude operator for the
field, measured in photon-number units; the expectation value of the
field, $\langle E(t)\rangle=\mbox{Re}(\langle a\rangle e^{-i\omega
t})$, oscillates with the amplitude and phase of $\langle a\rangle$.
The variance of $E$ characterizes the noise in the signal; for
phase-insensitive noise, for which $\langle(\Delta a)^2\rangle=0$
{(we use $\Delta O=O - \expt{O}$ here and throughout)}, this
variance is constant in time and given by
\begin{align}
2\langle(\Delta E)^2\rangle=\langle|\Delta a|^2\rangle
=\frac{1}{2}(\Delta x_1^2+\Delta x_2^2)
\ge\frac{1}{2}\,.
\label{eq:inputnoise}
\end{align}
Here $\langle|\Delta a|^2\rangle\equiv\frac{1}{2}\langle\Delta
a\Delta a^\dagger+\Delta a^\dagger\Delta a\rangle$ is the
symmetrically-ordered second moment of $a$.  The inequality follows
directly from the uncertainty principle for the quadrature
components, $\langle(\Delta x_1)^2\rangle\langle(\Delta
x_2)^2\rangle\ge1/4$. The lower bound is the half-quantum of
zero-point (or vacuum) noise and is saturated if and only if the mode
is in a coherent state~$|\alpha\rangle$.

The objective of phase-preserving linear amplification is to increase
the size of the input signal by a (real) amplitude gain $g$,
regardless of the input phase, while introducing as little noise as
possible.  The amplification of the input signal can be expressed as
a transformation of the expected complex amplitude,
\begin{equation}
\langle a_{\rm out}\rangle=g\langle a_{\rm in}\rangle\,.
\label{eq:expio}
\end{equation}
A {\em perfect linear amplifier\/} would perform this feat while
adding no noise; in the Heisenberg picture, the primary mode's
annihilation operator, not just its expectation value, would
transform from input to output~as
\begin{equation}
a_{\rm out}=ga_{\rm in}\,.
\label{eq:ioperfect}
\end{equation}
The second-moment noise would be amplified by the power gain $g^2$,
i.e., $\langle|\Delta a_{\rm out}|^2\rangle=g^2\langle|\Delta a_{\rm
in}|^2\rangle$.  The amplifier's output would be contaminated by the
same noise as the input, blown up by a factor of $g^2$, but the
amplification process would not add any noise to the amplified input
noise.

There are, however, no perfect phase-preserving linear amplifiers;
the transformation~(\ref{eq:ioperfect}) does not preserve the
canonical commutation relation and thus violates unitarity.
Physically, this is the statement that amplification of the primary
mode requires it to be coupled to other physical systems, not least
to provide the energy needed for amplification; these other systems,
which can thought of as the amplifier's internal degrees of freedom,
necessarily add noise to the output.  This physical requirement is
expressed in an input-output relation~\cite{Haus1962a,Caves1982a},
\begin{equation}
a_{\rm out}=ga_{\rm in}+L^\dagger\,,
\label{eq:io}
\end{equation}
where the \textit{added-noise operator\/} $L$ is a property of the
internal degrees of freedom.  One usually assumes that $\langle
L^\dagger\rangle=0$ so as to retain the expectation-value
transformation~(\ref{eq:expio}).  Preserving the canonical
commutation relation between input and output requires that
\begin{equation}
[L,L^\dagger]=g^2-1\,,
\label{eq:Lcomm}
\end{equation}
which implies an uncertainty principle for the added noise,
\begin{equation}
\langle|\Delta L|^2\rangle\ge\frac{1}{2}(g^2-1)\,.
\label{eq:addednoise}
\end{equation}

The amplifier must be prepared to receive any input in the primary
mode, without having any idea what that input is going to be.  This
places the restriction that the primary mode and the internal degrees
of freedom \textit{cannot\/} be correlated before amplification.  The
total output noise is then the sum of the amplified input noise and
the noise added by the internal degrees of freedom:
\begin{equation}
\langle|\Delta a_{\rm out}|^2\rangle
=g^2\langle|\Delta a_{\rm in}|^2\rangle+\langle|\Delta L|^2\rangle
\ge g^2-\textstyle{\frac{1}{2}}\,.
\end{equation}
The lower bound follows from the uncertainty
principles~(\ref{eq:inputnoise}) and~(\ref{eq:addednoise}).  An
amplifier that achieves the lower bound in Eq.~(\ref{eq:addednoise}),
thus adding the least amount of noise permitted by quantum mechanics,
is called an {\em ideal linear amplifier}.

\subsection{Ideal, perfect, and immaculate linear amplifiers}
\label{subsec:immac}

We can formulate a more general description of linear amplifiers by
using the formalism developed in Ref.~\cite{CavComJiaPan12}, where we
showed that for any phase-preserving linear amplifier, its action on
an input state $\rho$ of the primary mode can be represented by an
amplifier map
\begin{align}\label{eq:E}
\rho_{\rm out}=\sE(\rho)={\rm Tr}_{B}[S(r)\rho\otimes\sigma S^\dagger(r)]\,.
\end{align}
In this expression, $\sigma$ is the input state of a (perhaps
fictitious) ancillary mode $B$, which has annihilation and creation
operators $b$ and $b^\dagger$, and $S(r)=e^{r(ab-a^\dagger
b^\dagger)}$ is the two-mode squeeze operator.  The amplitude gain is
given by $g=\cosh r$, and the noise properties of the amplifier are
encoded in $\sigma$.   The main result of Ref.~\cite{CavComJiaPan12}
is that the amplifier map is physical, i.e., is completely positive,
if and only if $\sigma$ is a physical ancilla state.

The $P$ function of the output state can be written as a convolution
of the $P$ function of the input state with the $Q$ distribution
of~$\sigma$:
\begin{align}
P_{\rm  out}(\beta)=
\int d^2\alpha\,\frac{Q_\sigma\bigl[-(\beta^*-g\alpha^*)/\sqrt{g^2-1}\,\bigr]}{g^2-1}
P_{\rm in}(\alpha)\,.
\end{align}
We specialize for the remainder of this subsection to a
coherent-state input $|\alpha\rangle$, for which the input $P$
function is a $\delta$-distribution and the output $P$ function is
obtained by displacing and rescaling the $Q$ distribution of
$\sigma$,
\begin{align}
P_{\rm  out}(\beta)
=\frac{Q_\sigma\bigl[-(\beta^*-g\alpha^*)/\sqrt{g^2-1}\,\bigr]}{g^2-1}\,.
\end{align}
Moments of $\alpha$ calculated using the $P$ function give normally
ordered moments of $a$ and $a^\dagger$.

An ideal linear amplifier corresponds uniquely to the case where the
input ancilla state is vacuum, i.e., $\sigma = \op{0}{0}$, giving
rise to an output $P$ function
\begin{align}
P_{\rm out}(\beta)
=\frac{e^{-|\beta-g\alpha|^2/(g^2-1)}}{\pi(g^2-1)}\,.
\label{eq:Pout}
\end{align}
The displacement of the $Q$ distribution indicates that the input
complex amplitude is amplified as in Eq.~(\ref{eq:expio}), and the
rescaling of the $Q$ distribution confirms that the total (symmetric)
output noise is $\langle|\Delta a_{\rm out}|^2\rangle=\langle\Delta
a_{\rm out}^\dagger\Delta a_{\rm
out}\rangle+\frac{1}{2}=g^2-\frac{1}{2}$.

We can embed the ideal-amplifier map in a sequence of maps for both
physical and unphysical amplifiers by considering ancilla states of
thermal form,
\begin{align}
\sigma = \frac{1}{\mu^2}\!\left( 1 -\frac{1}{\mu^2} \right)^{a\dg a}
=\frac{1}{\mu^2}\sum_{n=0}^\infty\left(1-\frac{1}{\mu^2}\right)^n \ket{n}\bra{n}
\,.
\label{eq:sigmamu}
\end{align}
When $\mu^2\in[1,\infty)$, $\sigma$ is a physical thermal state, with
dimensionless inverse temperature $\beta$ given by
$\mu^2=(1-e^{-\beta})^{-1}$; $\mu^2=1$ gives the vacuum state.  When
$\mu^2\in[0,1)$, however, $\sigma$ has negative eigenvalues and thus
is unphysical.  When $\mu^2\in(\frac{1}{2},\infty)$, the trace of
$\sigma$ is well defined and equal to 1, but when $\mu^2\in[0,1/2]$,
the series for the trace of $\sigma$ diverges; $\mu^2=1/2$ makes
$\sigma$ the parity operator.  The amplifier maps corresponding to
unphysical $\sigma$ are not completely positive and thus are
unphysical~\cite{CavComJiaPan12}.  In the following, we sometimes use
quotes to warn the reader that $\sigma$ might not be physical.

\begin{figure}[htbp]
\includegraphics[width=0.45\textwidth]{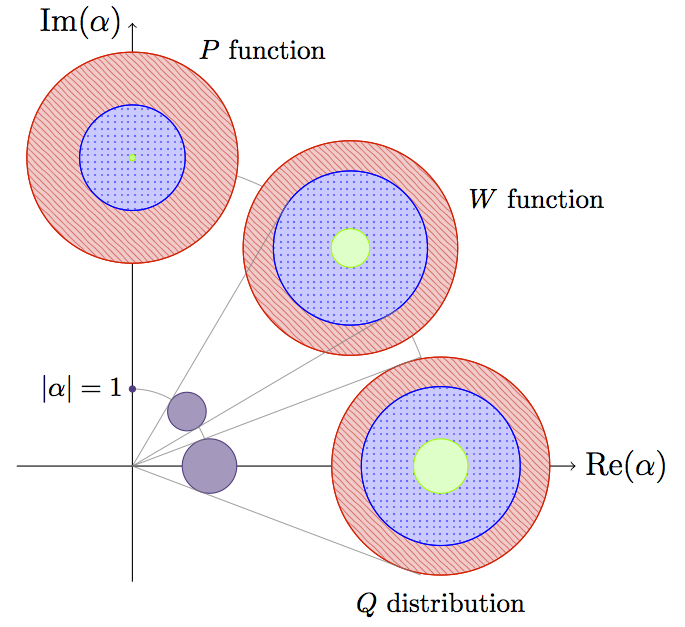}
\caption{(Color online) Ball-and-stick phase-space depictions of
input and output noise for ideal ($\mu^2=1$), perfect
($\mu^2=\frac12$), and immaculate ($\mu^2=0$) amplifiers defined by
the amplifier map~(\ref{eq:E}) with initial ancilla
``state''~(\ref{eq:sigmamu}).  Color and fill conventions: solid
(purple) fill is used for input noise; (red) fill with slanted lines for
the output noise of an ideal amplifier; (blue) fill with dots for the
output of a perfect amplifier; and solid (green) fill for the output of
an immaculate amplifier.  The primary-mode input is a coherent state
$|\alpha\rangle$ with $|\alpha|=1$, and the gain is $g=4$, giving the
output state a mean that lies on a circle of radius $g|\alpha|=4$.
The input and output states are represented by noise circles centered
at the mean complex amplitude (the stick) and having radius
$\Sigma/2\sqrt2$ (the ball), where
$\Sigma^2=\langle|\Delta\alpha|^2\rangle$ is the variance of the
complex amplitude calculated from the appropriate quasidistribution:
for the normal ordering of the $P$ function,
$\Sigma_P^2=\langle\Delta a^\dagger\Delta a\rangle$; for the
symmetric ordering of the Wigner~$W$ function,
$\Sigma_W^2=\frac{1}{2}(\langle\Delta a^\dagger\Delta
a\rangle+\langle\Delta a\Delta
a^\dagger\rangle)=\Sigma_P^2+\frac{1}{2}$; for the antinormal
ordering of the $Q$~distribution, $\Sigma_Q^2=\langle\Delta a\Delta
a^\dagger\rangle=\Sigma_W^2+\frac{1}{2}$.  The $P$-function depiction
is the one suggested by the amplifier map~(\ref{eq:E}): the dot
($\Sigma_P=0$) for the input coherent state $|\alpha\rangle$ is
amplified by an immaculate amplifier to a dot for the output coherent
state $|g\alpha\rangle$; the output for a perfect amplifier has
additional noise $\Sigma_P^2=\frac{1}{2}(g^2-1)$, and the output for
an ideal amplifier has additional noise $\Sigma_P^2=g^2-1$. The
symmetrically ordered moments of the Wigner $W$ function give the
traditional picture of amplifier noise: the input coherent state,
represented by a circle corresponding to $\Sigma_W^2=\frac{1}{2}$,
has its noise amplified by a perfect amplifier along the (grey)
radial lines to the circle with $\Sigma_W^2=\frac{1}{2}g^2$; the
output of an ideal amplifier has additional noise
$\frac{1}{2}(g^2-1)$, giving total noise
$\Sigma_W^2=g^2-\frac{1}{2}$, and the output of an immaculate
amplifier has its noise reduced by $\frac{1}{2}(g^2-1)$ to the
coherent-state value $\Sigma_W^2=\frac{1}{2}$.  The antinormally
ordered moments of the Husimi $Q$ distribution give a picture suited
to discussion of simultaneous measurements of the quadrature
components (see text): the input coherent state, represented by a
circle corresponding to $\Sigma_Q^2=1$, has its noise amplified by an
ideal amplifier along the (grey) radial lines to a circle with
$\Sigma_Q^2=g^2$; the output of a perfect amplifier has less noise by
$\frac{1}{2}(g^2-1)$, giving total noise
$\Sigma_Q^2=\frac{1}{2}(g^2+1)$, and the output of an immaculate
amplifier has its noise reduced by $g^2-1$ to the coherent-state
value $\Sigma_Q^2=1$.
}\label{fig1}
\end{figure}

The $Q$ function for~$\sigma$,
$Q_\sigma(\alpha)=e^{-|\alpha|^2/\mu^2}/\pi\mu^2$, is well behaved on
the entire range $\mu^2\in(0,\infty)$ and becomes a $\delta$-function
when $\mu^2=0$.  The output $P$ function is the Gaussian
\begin{align}
P_{\rm out}(\beta,\mu^2)
=\frac{1}{\pi\mu^2(g^2-1)}
e^{-|\beta-g\alpha|^2/\mu^2(g^2-1)}\,,
\label{eq:Pnuout}
\end{align}
which has normally ordered output noise $\langle\Delta
a^\dagger\Delta a\rangle=\mu^2(g^2-1)$ and, hence,
symmetrically-ordered output noise~\cite{sordering}
\begin{align}
\langle|\Delta a_{\rm out}|^2\rangle
=\langle\Delta a_{\rm out}^\dagger\Delta a_{\rm out}\rangle+\textstyle{\frac{1}{2}}
=\mu^2(g^2 - 1)+\textstyle{\frac{1}{2}}.
\end{align}
The output $Q$ distribution is
\begin{equation}
Q_{\rm out}(\beta,\mu^2)
=\frac{1}{\pi[\mu^2(g^2-1)+1]}
e^{-|\beta-g\alpha|^2/[\mu^2(g^2-1)+1]}\,.
\label{eq:Qnuout}
\end{equation}

We now focus on three amplifiers of interest, which correspond to
three values of $\mu^2$:
\begin{enumerate}
\begin{item}
The {\em ideal linear amplifier} (physical), which corresponds to
$\mu^2=1$ and which adds the minimal amount of (symmetrically
ordered) noise permitted by quantum mechanics.
\end{item}
\begin{item}
The {\em perfect linear amplifier} (unphysical), $\mu^2=1/2$,
whose (symmetrically ordered) output noise consists only of the
amplified input noise.
\end{item}
\begin{item}
The unphysical $\mu^2=0$ amplifier, which we christen the {\em
immaculate linear amplifier}, because it is better than perfect,
and which takes an input coherent state $|\alpha\rangle$ to an
amplified output coherent state $|g\alpha\rangle$.  We let $\sA$
denote the amplifier map~(\ref{eq:E}) for the case of an
immaculate linear amplifier, i.e.,
\begin{align}
\sA(|\alpha\rangle\langle\alpha|)
=|g\alpha\rangle\langle g\alpha|\,.
\label{eq:A}
\end{align}
\end{item}
\end{enumerate}

The operation of these three amplifiers can be understood intuitively
in terms of how the output noise arises from amplified input noise
and added noise.  The three canonical quasidistributions, the $P$
function, the Wigner~$W$ function, and the Husimi~$Q$
distribution~\cite{GarrisonChiao2008}, with their different operator
orderings, quantify the noise differently and thus provide three
different perspectives on the relation between input and output
noise. In \frf{fig1} we illustrate the amplification transformations
for ideal, perfect, and immaculate amplifiers.  The transformations
can be summarized in terms of ball-and-stick phase-space diagrams
that depict the input and output noise as circles of uncertainty
centered at the input and output mean complex amplitudes.  We give
such diagrams for the normally ordered variances corresponding to
input and output $P$~functions, as in Eq.~(\ref{eq:Pnuout}), and also
for the symmetrically ordered moments of input and output Wigner~$W$
functions and the antinormally ordered moments of input and output
Husimi $Q$ distributions.

The $P$-function perspective, with its normally ordered moments, is
matched to {the immaculate amplifier map~(\ref{eq:A})}. The
immaculate amplifier takes an input coherent state to an amplified
coherent state; in the $P$-function depiction, it takes an input dot
in the phase plane to an output dot, without adding any noise.  All
the output noise for a perfect or an ideal amplifier appears to be
added noise.

The symmetrically ordered moments of the Wigner function give the
traditional perspective on amplifier noise.  A perfect amplifier
amplifies input coherent-state noise without adding any noise.  An
ideal amplifier adds further noise $\langle|\Delta
L|^2\rangle=\frac{1}{2}(g^2-1)$, and an immaculate amplifier
subtracts the same amount of noise.

The antinormally ordered moments of the $Q$ function give a picture
matched to an ideal amplifier.  The input noise of a coherent state
is amplified by an ideal amplifier to produce the output noise
without addition of any further noise.  A perfect amplifier has less
output noise by $\frac{1}{2}(g^2-1)$, and an immaculate amplifier
less noise by $g^2-1$.

\subsection{Na\"\i ve uncertainty-principle bounds on probabilistic $\mu^2$ amplifiers}
\label{subsec:upbounds}

The antinormally ordered noise of the $Q$ function has a physical
interpretation that sheds light on the performance of linear
amplifiers.  Suppose one wishes to determine the center of a coherent
state by making simultaneous measurements of the two quadrature
components.  The statistics of ideal simultaneous measurements are
given by the $Q$ distribution~\cite{Arthurs1965a}, so in $\nu$ such
measurements, one can determine the center with uncertainty $(\delta
x_1)_{\rm in}/\sqrt\nu=(\delta x_2)_{\rm in}/\sqrt\nu=1/\sqrt\nu$;
the uncertainties here, distinguished by a $\delta$, are calculated
from the $Q$ distribution, i.e., using antinormal ordering.
Alternatively, one could amplify the coherent state with an ideal
linear amplifier and determine the center of the output state with
uncertainty $(\delta x_1)_{\rm out}/\sqrt\nu=(\delta x_2)_{\rm
out}/\sqrt\nu=g/\sqrt\nu$; this allows one to determine the center of
the input coherent state with the same uncertainty as measurements at
the input, i.e., $(\delta x_1)_{\rm out}/g\sqrt\nu=(\delta x_2)_{\rm
out}/g\sqrt\nu=1/\sqrt\nu$.  The point of linear amplification is to
make a signal much larger so it can be detected by less sensitive
measurements.  That it is possible to determine the input with
exactly the same sensitivity by measuring either the input or the
output is an alternative way of characterizing the performance of an
ideal amplifier.

It is interesting to apply this sort of thinking to the unphysical
amplifiers with $\mu^2<1$; if one could construct such an amplifier,
one could determine the center of an input coherent state with
uncertainty
\begin{align}
\frac{(\delta x_1)_{\rm out}}{g\sqrt\nu}
=\frac{(\delta x_2)_{\rm out}}{g\sqrt\nu}
=\frac{\sqrt{\mu^2(g^2-1)+1}}{g\sqrt\nu}\,.
\end{align}
This violates the uncertainty-principle bound for any $\mu^2<1$ and
thus provides another way of seeing why the amplifiers with $\mu^2<1$
are unphysical.

A potential way to make such an amplifier physical is to make it
nondeterministic, so that it only works with probability $\Pcheck$.
Then, since only $\Pcheck\nu$ of the trials are effective, one can
determine the center of the input coherent state with uncertainty
$(\delta x_1)_{\rm out}/g\sqrt{\Pcheck\nu}=(\delta x_2)_{\rm
out}/g\sqrt{\Pcheck\nu}$.  Requiring that this uncertainty not best
the uncertainty-principle bound,
\begin{align}
\frac{(\delta x_1)^2_{\rm out}}{\Pcheck g^2}
=\frac{(\delta x_2)^2_{\rm out}}{\Pcheck g^2}\ge1\,,
\label{eq:up}
\end{align}
gives us a bound on the working probability,
\begin{align}
\Pcheck
\le\frac{(\delta x_1)^2_{\rm out}}{g^2}
=\frac{(\delta x_2)^2_{\rm out}}{g^2}
=\mu^2+\frac{1-\mu^2}{g^2}\,.
\label{eq:Pmu}
\end{align}

{Another way to express the bound~(\ref{eq:Pmu}) is in terms
of the root-probability--SNR product, $\sqrt{\Pcheck}{\rm SNR}$,
where if $x_1$ and $x_2$ represent the amplitude and phase
quadratures ($\langle x_1\rangle=\sqrt2|\alpha|$ and $\langle
x_2\rangle=0$), the signal-to-noise ratio is defined as ${\rm
SNR}\equiv\langle x_1\rangle/\delta x_1=\langle x_1\rangle/\delta
x_2$.  The root-probability--SNR product is a measure of the
resolvability of states.  The uncertainty-principle
bound~(\ref{eq:Pmu}) on success probability is equivalent to the
requirement that amplification not increase this resolvability, i.e.,
\begin{align}
\sqrt{\Pcheck}\,{\rm SNR}_{\rm out}\le{\rm SNR}_{\rm in}=\sqrt2|\alpha|\,.
\label{eq:SNR_bound}
\end{align}
The root-probability--SNR product provides the same information as
the uncertainty-principle bound, but without referring output
quantities to the input.  We consider the root-probability--SNR
product again in Sec.~\ref{sec:kraus}.}

It is worth noting that since the output state $\rho_{\rm out}$ is
Gaussian, its fidelity with $|g\alpha\rangle$ is the inverse of the
antinormally-ordered output variances:
\begin{equation}
F(\mu^2)=\langle g\alpha|\rho_{\rm out}|g\alpha\rangle
=\pi Q_{\rm out}(g\alpha)=\frac{1}{\mu^2(g^2-1)+1}\,.
\end{equation}
This gives a bound on the probability-fidelity product,
\begin{equation}
\Pcheck(\mu^2)F(\mu^2)\le\frac{1}{g^2}\,,
\label{eq:PFP}
\end{equation}
which is independent of $\mu^2$ and achieved by an ideal linear
amplifier.  The probability-fidelity product can be regarded as the
overall probability to reach the target state $|g\alpha\rangle$. Such
products appear again throughout our analysis.

For the remainder of the paper, we focus on the immaculate linear
amplifier ($\mu^2=0$), for which the probability bound~(\ref{eq:Pmu})
becomes $\Pcheck\le1/g^2$. Our analysis shows that a nondeterministic
immaculate linear amplifier only works with high fidelity on a
portion of phase space, where it has considerably less chance of
working than this bound.  It thus does considerably worse than a
deterministic linear amplifier in determining the center of an input
coherent state.  This suggests that such devices should not be
thought of primarily as linear amplifiers.  They could be used,
however, as probabilistic, approximate cloners, a task that we
consider now.

\section{Amplifiers and cloning}
\label{sec:relation_cloning}

Exact, deterministic cloning is not allowed by quantum
mechanics~\cite{WooZur82,Dieks82,Yuen86}.  For coherent states, the
impossibility of exact, deterministic cloning corresponds to the
impossibility of deterministic immaculate amplification.  If one has
$M$ clones of a coherent state $|\alpha\rangle$, they can be
coherently combined in an $M$-port device to produce $M-1$ vacuum
states and a single amplified coherent state $|g\alpha\rangle$, with
$g=\sqrt M$; running an amplified coherent state $|g\alpha\rangle$
backwards through the same device splits that state into $M$ clones.
This equivalence between cloning and immaculate amplification is the
basis for links between cloning and amplification (see, e.g.,
Refs.~\cite{Her82,WooZur82}); here we summarize the links and the
terminology relevant to {this
paper~\cite{CerfFiura06,BraunsteinvanLoock05}.}

The cloning literature phrases the task of cloning in terms of
transforming $N$ replicas of the state to be cloned into some number
$M$ of identical clones; this is termed ``$N$ to $M$'' cloning and is
often denoted $N\rightarrow M$.  An amplifier with amplitude gain $g$
can be thought of as doing $1\rightarrow M=\sqrt g$ cloning.  Since
exact, deterministic cloning is ruled out by the no-cloning theorem
when $M>N$, one must drop either exactness, considering instead {\em
noisy\/} or {\em approximate\/} cloning~\cite{BuzHill96}, or
determinism, considering instead {\em probabilistic\/} cloning.

Consider first approximate, deterministic cloning.  The standard
measure of performance for approximate cloning is the fidelity $F$ of
the clones with the desired target state.  If the clones all have the
same fidelity with the target state, the cloning process is said to
be {\em symmetric}.  If the fidelity of the clones is independent of
the input state, the cloning is called {\em universal}.

It is known~\cite{Lindblad00,CerfFiura06} that the optimal fidelity
for cloning coherent states $\ket{\alpha}$ to $M$ clones that have
Gaussian noise is achieved by using an ideal linear amplifier with
gain $g=\sqrt M$, followed by an $M$-port device that splits the
amplified state into $M$ approximate clones, each of which has the
marginal state $\rho_\alpha$.  The state $\rho_\alpha$ has $P$
function $P_\alpha(\beta)=g^2P_{\rm out}(g\beta)$ [see
Eq.~(\ref{eq:Pout})], and the corresponding $Q$ distribution is
\begin{align}
Q_\alpha(\beta)
=\frac{e^{-|\beta-\alpha|^2/(2-1/g^2)}}{\pi(2-1/g^2)}\,.
\end{align}
The output fidelity,
\begin{align}
F_{1\rightarrow M}=\langle\alpha|\rho_\alpha|\alpha\rangle=
\pi Q_\alpha(\alpha)=\frac{M}{2M-1}\,,
\end{align}
is a function of the gain alone, independent of the amplitude of the
input state~\cite{BraBuzHil01,CerfFiura06}.  This output fidelity
limits to $\frac12$ as $M\rightarrow\infty$.

Suppose instead that one desires perfect clones and is thus willing
to put aside determinism.  This is called exact ($F=1$),
probabilistic cloning~\cite{DuaGuo98}, and the appropriate measure of
performance is the probability $\Pcheck$ that the cloning process
works. In probabilistic cloning, one usually restricts to a finite
set of input states and attempts to clone these states optimally. The
restriction on input states is referred to as {\em state-dependent\/}
cloning.

In \srf{sec:usd}, we consider exact, but probabilistic immaculate
amplification. Given the equivalence between immaculate amplification
and exact cloning, this can equally well be thought of as exact,
probabilistic, $1\rightarrow M=\sqrt g$ cloning of coherent states.
We show that exact, probabilistic immaculate amplification of all
coherent states---or even of all the coherent states on a circle
centered at the origin of phase space---cannot occur with a nonzero
probability of success.  If, however, the input coherent states are
restricted to a finite set equally spaced around a circle centered at
the origin, exact immaculate amplification can occur with a success
probability given by the probability of unambiguously discriminating
the input coherent states~\cite{Che98,CheBar98}. Once one has
identified unambiguously the input state, one can do any state
transformation, including making an amplified coherent state or
making as many exact clones as one wants.  Thus we have a recipe for
making an exact, probabilistic immaculate amplifier or an exact,
probabilistic, state-dependent cloner.

In \srf{sec:kraus}, we derive rigorous bounds on the success
probability of an amplifier that amplifies coherent states near the
origin immaculately with fidelity near unity, but has output fidelity
that decreases to zero as the amplitude of the input coherent states
increases.  Since the output states do not have Gaussian noise, the
connection to cloning is not precise, but for coherent states near
the origin, these amplifiers can be thought of as cloners that are
approximate, probabilistic, and state dependent.

There is some cloning literature that considers various combinations
of approximate, probabilistic, and state-dependent cloning.  For
example, Chefles and Barnett~\cite{CheBar99a} interpolate between
exact, probabilistic, state-dependent cloners and approximate,
deterministic cloners, including both fidelity and success
probability as performance measures, but only for two input states, a
restriction that makes their results too limited for our purposes.
There is also work on cloning for a distribution of input coherent
states~\cite{CocRalDol04}, which derives the optimal average fidelity
of a $1\rightarrow2$ cloner that acts on a Gaussian distribution with
width $\Delta$ centered at the origin.  As the width goes to zero,
the average fidelity not surprisingly approaches unity.

\section{Prior work on probabilistic immaculate amplification}
\label{sec:paprior}

Ralph and Lund~\cite{RalLu09} conceived the notion of an immaculate
linear amplifier and proposed a probabilistic implementation (what
they called a nondeterministic, noiseless linear amplifier) described
by a quantum operation
\begin{align}\label{eq:rallun}
\sE_{\rm amp} (\rho) = \Echeck(\rho)+\sE_{\rm fail} (\rho),
\end{align}
where $\Echeck$ is the quantum operation when the amplifier works and
$\sE_{\rm fail}$, the quantum operation when it fails, describes its
fallible nature.

Ralph and Lund~\cite{RalLu09} and collaborators~\cite{XiaRalLun10}
suggested that the most straightforward incarnation of a
probabilistic immaculate amplifier is to have
\begin{align}\label{eq:Echeck}
\Echeck(\op{\alpha}{\alpha}) = \Pcheck\!\op{g\alpha}{g\alpha}
\end{align}
for all input coherent states, where $\Pcheck$ is the
state-independent probability that the amplifier works.  Since this
makes $\Echeck=\Pcheck\sA$, i.e., a multiple of the map~(\ref{eq:A}) for a deterministic immaculate amplifier, it is not
completely positive unless the success probability is zero. Indeed,
quite generally, if $\Echeck$ works as a linear amplifier with
uniform success probability over the entire phase plane, complete
positivity imposes the same restrictions on $\Echeck$ as for a
deterministic linear amplifier; in particular, $\sE_{\rm amp}$ would
be just as noisy as a deterministic amplifier, the only difference
being that some of the time the amplifier wouldn't work at all.  To
make an immaculate amplifier physical, one must make it not just
probabilistic, but also drop the idea that it can work immaculately
over the entire phase plane with uniform success probability.  In
making models of immaculate amplification, this is precisely what
Ralph and Lund~\cite{RalLu09} and Fiur\'a\v{s}ek~\cite{Fiur04a} did.

For the remainder of this section, we review some of the theoretical
proposals for and experimental realizations of \erf{eq:rallun}.  Here
implementation is interpreted as meaning that the amplifier works
immaculately with high fidelity in a restricted region of phase space
near the origin and with the success probability $\Pcheck$ depending
on the distance of the input coherent state from the origin.

{\em Quantum-scissors proposal.} Ralph and Lund originally proposed
to implement \erf{eq:Echeck}\ using a network of beam splitters,
single-photon sources, and single-photon detectors, as illustrated in
\frf{fig2}.  An input coherent state $|\alpha\rangle$ is split up
equally at an $N$-port splitter, each output $|\alpha/\sqrt N\rangle$
is processed through a modified ``quantum scissors''
(MQS)~\cite{PegPhiBar98}, and the outputs of the quantum scissors are
recombined at a second $N$-port splitter.  Successful immaculate
amplification requires heralding on the MQSs so that they work
correctly and on vacuum detection in $N-1$ outputs of the second
splitter. These heralding requirements mean that quantum-scissors
proposal is probabilistic, and its region of high-fidelity immaculate
amplification is restricted by the requirement that $|\alpha|^2\ll
N$.  Even within this phase-plane region, the fidelity with the
target state $|g\alpha\rangle$ is a function of the amplitude
$|\alpha|$ of the input coherent state.

\begin{figure}[htbp]
\includegraphics[width=0.47\textwidth]{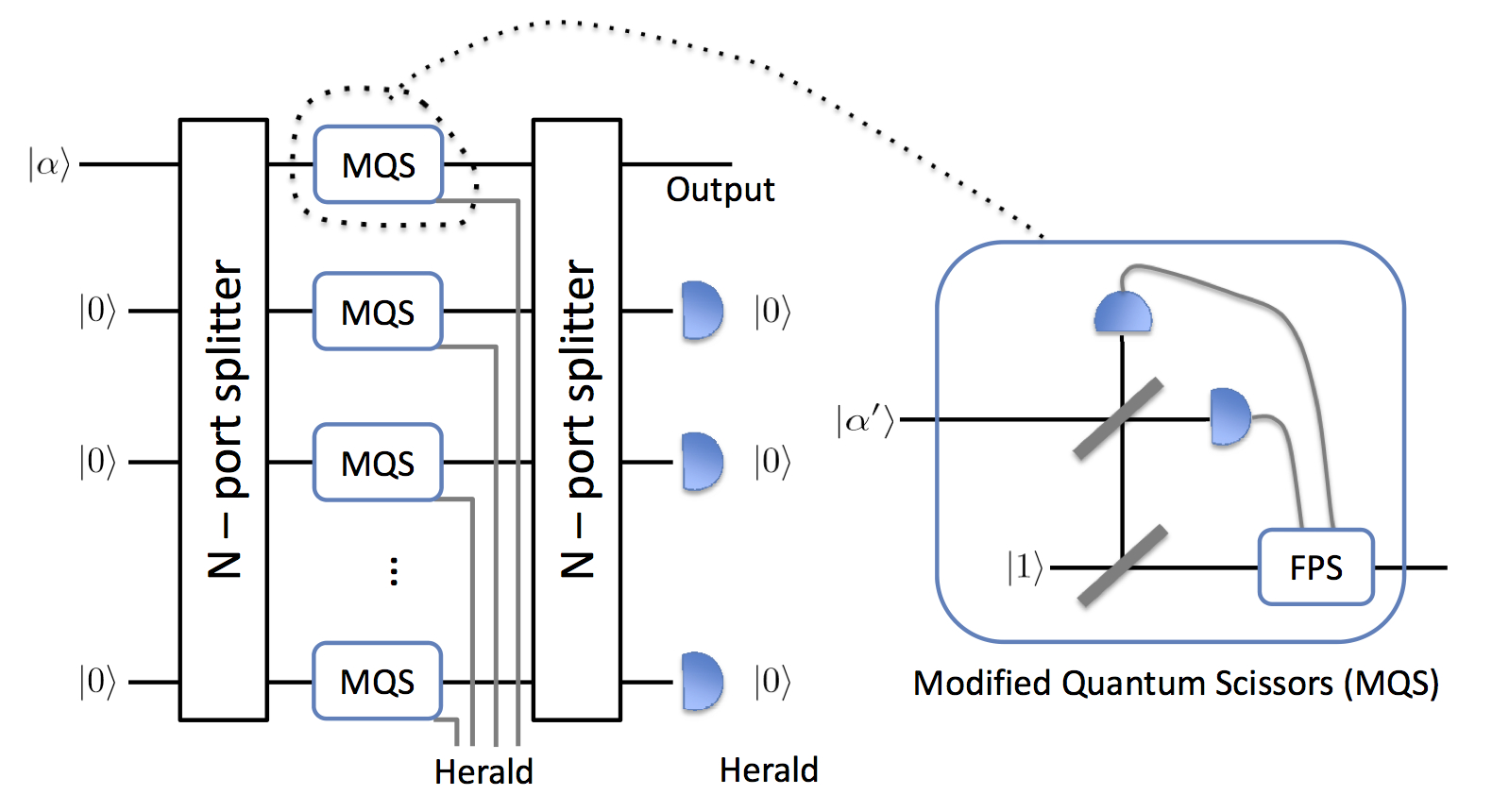}
\caption{Device that approximates an immaculate amplifier (figure
based on Fig.~1 of~\cite{XiaRalLun10}).  An incident coherent state
is split equally into $N$ modes at an $N$-port splitter.  The state
of each mode is a coherent state $|\alpha'\rangle$, where
$\alpha'=\alpha/\sqrt N$; $N$ is chosen large enough that $\alpha'=
\alpha/\sqrt{N}\ll1$, so that $\ket{\alpha'}=\ket{0} + \alpha'\ket{1}
+ O(|\alpha'|^2)$ is well approximated by its vacuum and one-photon
pieces. Each of the $N$ modes enters a modified ``quantum scissors''
(MQS)~\cite{PegPhiBar98}, shown on the right, which is the heart of
the amplifier.  When the two detectors in the MQS get results 1,0 or
0,1, the MQS is said to work; a feedforward phase shift (FPS) by
$\pi$, controlled on one of the two outcomes, is applied to the
device's output mode.  The result of these manipulations is that,
conditioned on the MQS working, it implements the transformation
$\ket{\alpha'}\rightarrow(1 +
g\alpha'a^\dag)\ket{0}=\ket{g\alpha'}_{\rm trunc}$, i.e., truncation
of the state to the vacuum--one-photon sector and change of the
relative weights of the vacuum and one-photon contributions so that
the one-photon weight is increased; the gain is determined by the
transmissivities and reflectivities of the beamsplitters in the MQS.
The amplified and truncated states, $\ket{g\alpha'}_{\rm trunc}$, are
recombined at a second $N$-port splitter. Conditional on detecting
vacuum in $N-1$ outputs of this splitter, the output mode is in the
amplified state $\ket{g\alpha}$ in the limit that $N\rightarrow
\infty$.  Successful immaculate amplification thus corresponds to
heralding on the desired outcome of all of the MQSs, as well as
vacuum detection in the $N-1$ ports of the final $N$-port splitter.}
\label{fig2}
\end{figure}

In Ref.~\cite{Jeffers10}, Jeffers tried to reduce the need to make
$N$ so large by constructing a quantum-scissors device that works at
the two-photon level, i.e., that implements the
truncation-and-amplification transformation
$|\alpha'\rangle=c_0\ket0+c_1\ket1+c_2\ket2+O(|\alpha'|^3)\rightarrow
c_0\ket{0}+gc_1\ket{1}+g^{2}c_2\ket{2}$. Though this is a nice idea,
there is a catch: it requires lossy beam splitters or a beamtritter.
Numerically it was shown that, for $|\alpha|^2=0.1$, a single
two-photon device performs better than $N=3$ single-photon MQSs with
respect to the fidelity of the output with the target amplified state
$\ket{g \alpha}$ and the success probability.  No mention is made in
either Jeffers's or Ralph and Lund's work of how close these
implementations are to limits imposed by quantum theory.

{\em Quantum-scissors implementations.} One-photon scissors devices
have been implemented experimentally by Xiang {\em et
al.}~\cite{XiaRalLun10} and Ferreyrol {\em et
al.}~\cite{FerBarBla10,FerBlaBar11}.

The experiment by Xiang {\em et al.}~\cite{XiaRalLun10} used an
attenuated spontaneous parametric down-conversion source to produce
an input state $\rho_{\rm in} = (1-|\alpha|^2)\op{0}{0} +
|\alpha|^2\op{1}{1}$ where $|\alpha|^2\in [10^{-3},10^{-1} ]$.  This
state is an approximation to a uniform mixture of coherent states of
fixed amplitude; the motivation for considering this input state was to investigate the action of the amplifier on all states in
the mixture simultaneously.  As the value of $|\alpha|$ was so small,
i.e., $\alpha = \alpha'$, only $N=1$ quantum-scissors device is
needed. The domain of gains used in the experiment was $g\in
[\sqrt{2},2]$. For $g=\sqrt{3}$, the experimental data showed that
the amplifier was linear over the range $|\alpha| \in
[10^{-3},2\times10^{-2}]$.

Ferreyrol {\em et al.}~\cite{FerBarBla10,FerBlaBar11} implemented
quantum-scissors-type amplifiers with $N=1$, input coherent states
with $|\alpha| \in [5.5\times 10^{-2},1]$, and $g\in [0.25, 2]$.
Their theoretical modeling and experimental results are in agreement
with the modeling and results in Ref.~\cite{XiaRalLun10}. The first
data point is in the region of phase space where the device has
linear gain.  Very quickly, however, the gain decreases for input
states with $|\alpha|>5.5\times10^{-2}$. Their data also show that as
the coherent-state amplitude increases, the probability of the
amplifier's working increases, and the output state is increasingly
distorted away from the target coherent state.  {These
behaviors appear in our analysis of quantum limits on immaculate
amplifiers in \srf{sec:kraus}.}

{\em Photon addition and subtraction proposals.}
Fiur\'a\v{s}ek~\cite{Fiurasek09} and, separately, Marek and
Filip~\cite{MarFil10} attempt to approximate the transformation in
\erf{eq:Echeck} by adding and then subtracting $M$ photons from a
low-amplitude coherent state.  The transformation for $M=1$ is
$aa\dg( \ket{0} +\alpha\ket{1})\rightarrow a( \ket{1}
+\sqrt{2\alpha}\ket{2})\rightarrow  \ket{0} +2\alpha \ket{1}$, which
has a gain of~$2$.  This will not act like a linear amplifier unless
$|\alpha|\alt1$.  Generalizing to $M$-photon addition and
subtraction, the gain becomes $g=M+1$.  The chief problem with this
method is the experimental infeasibility of $M$-photon addition and
subtraction for $M$ more than a very few.

{\em Photon addition and subtraction implementations.}  Zavatta {\em
et al.}~\cite{ZavFiuBel11} reported an experimental implementation of
a single-photon addition and subtraction device ($M=1$), which had
$|\alpha| \in [0.2,1]$, and $g\in [1.25, 2]$.  For input
$|\alpha|>0.5$ the fidelity of the output state with $\ket{g\alpha}$
dropped dramatically, and the appearance of the output Wigner
function departed noticeably from the target Wigner function in a way
we return to in \srf{sec:kraus}.   The authors point out that an
equivalent quantum-scissors device performs worse with respect to
gain and fidelity, both of which decrease quicker with increasing
$|\alpha|$ in the scissors case.

{\em Proposals for noise addition followed by photon subtraction.} To
overcome the difficulties of adding $M$ photons, Marek and
Filip~\cite{MarFil10} suggested one could simply add
phase-insensitive noise (random displacements on the phase plane) and
then do $M$-photon subtraction.  Intuitively this can be understood
as follows: adding noise increases the phase space area of the state;
the subsequent photon subtraction enhances the larger photon numbers,
producing an amplified final state that is, roughly speaking,
squeezed in the amplitude direction.  An explicit formula is given
for the success rate as a function of the input coherent state, $M$,
and the mean number of thermal photons added.

{\em Implementation of noise addition followed by photon
subtraction.} Usuga {\em et al.}~\cite{UsuMulWit10} and
Usuga~\cite{UsugaThesis12} describe the preparation of a displaced
thermal state which is intended to correspond to a coherent state
with added thermal noise. The parameters used in their experiments
are $|\alpha| = 0.431$, $g\in [1,2]$, and $M\in [1,4]$.  For $g>2$
($M>1$), the authors found the probability of success decreased
drastically, and the state started to deform (also see Ref.~\cite{MulWitMar12}).

{\em Discussion.} From the theory and experiments summarized above,
several conclusions can be drawn.  First, all of the devices produce
an output state with high fidelity to the target coherent state
$\ket{g\alpha}$ only over a restricted region of the phase plane
centered on the origin.  Second, although the theoretical proposals
allow for high gains and high input amplitudes, current
implementations are restricted to small gains $g\alt2$ and small
input amplitudes $|\alpha|\alt2$ by technical limitations.  Third,
even for these small gains and small input amplitudes, these devices
fail almost all of the time.

Most previous work on this subject has focused on proposing and
analyzing the performance of specific schemes for probabilistic
immaculate amplification.  We take a different tack: {we
provide a general} analysis of the performance of any device that
attempts to approximate immaculate linear amplification.  We
characterize the amplifier by its gain and the region of the phase
plane over which it operates with high fidelity, and we derive
fundamental quantum limits on the probability that the amplifier
works.

\section{USD bounds on probabilistic immaculate amplification}
\label{sec:usd}

Quantum state discrimination is a decision-theoretic task in which an
agent, who has the ability to perform any measurement he wishes, is
handed a single state drawn from a known set of states and is told to
determine which of the states he received.  Our chief interest here
is unambiguous state discrimination (USD): the agent is told never to
misidentify the state, at the cost of sure and sudden death, but is
allowed throw up his hands in despair and refuse to make a decision.
A set of states can be discriminated unambiguously if and only if
they are linearly independent~\cite{Che98}; there is a nonzero
probability for no decision unless the states are orthogonal.  In
this section we apply USD bounds to the performance of exact
immaculate amplifiers.  We use the USD formalism in two ways.

The first is to provide upper bounds on the working probability of an
immaculate amplifier. Let $\wp(\checkmark)$ be the probability that
an immaculate amplifier works exactly on a set of input coherent
states. Suppose that $P^{\rm B}$ is the optimal probability for
discriminating the input states and $P^{\rm A}$ is the corresponding
optimal probability for discriminating the amplified states.  The
amplified states, being further apart on the phase plane than the
input states, are easier to distinguish, so $P^{\rm A}>P^{\rm B}$.
The overall probability of successfully discriminating the amplified
states is $\wp(\checkmark)P^{\rm A}$.  Since $P^{\rm B}$ is optimal,
the amplification process cannot increase the distinguishability of
the states, so we must have $P^{\rm B}\ge\wp(\checkmark)P^{\rm A}$.
The result is a strict upper bound, $\wp(\checkmark)\le P^{\rm
B}/P^{\rm A}$, on the probability that the immaculate amplifier
works; we cannot warrant, however, that this upper bound can be
achieved.

The second way we use the USD formalism is to construct models of
immaculate amplifiers that have an achievable working probability.
Once one has used USD to identify one of the input states, one can
perform any unitary transformation on that state.  This procedure
always produces the right transformed state when it makes a decision;
consequently we call it, somewhat cumbersomely, an {\em exact,
finite-state, probabilistic state transformation}.  The
transformation could be the displacement of a coherent state required
to amplify it.  Since the optimal USD discrimination probability
$P^{\rm B}$ can be achieved in principle, the result is a model for
an immaculate amplifier that works with probability $P^{\rm B}$ on a
finite set of input coherent states.  We call such a model a {\em
finite-state, probabilistic immaculate amplifier}.

We note this formulation and subsequent analysis is similar to the
analysis performed by Dunjko and Andersson in Ref.~\cite{DunAnd12}.
Their results are not explicit about the dependence of the success
probabilities on gain and input amplitude, whereas we are.

\subsection{Helstrom bound for two coherent states}

Before turning to USD bounds on immaculate amplifiers, we consider a
related bound provided by the {minimal} error probability in
discriminating two nonorthogonal states.  Consider two coherent
states, $\ket{\alpha}$ and $\ket{\beta}$.  A measurement that
minimizes the chance of incorrectly identifying the state is known as
a Helstrom discrimination measurement~\cite{Hel76,WisMilBook}.  The
probability of successful identification is
\begin{align}
P_{\rm Hel}^{\rm B}(\checkmark)
&= \frac12 \! \Big ( 1 + \sqrt{1- |\ip{\beta}{\!\alpha}|^2} \Big)\nonumber\\
&= \frac12 \! \left ( 1 + \sqrt{1- e^{-|\alpha-\beta|^2}} \right),
\end{align}
where the superscript ``B'' reminds us that this probability is {\em
before\/} immaculate amplification.  It is apparent that as the
separation, $|\alpha-\beta|$, between the two states grows, the
states become orthogonal, and the probability of successful
discrimination approaches unity. In contrast, when
$|\alpha-\beta|\rightarrow0$, the success probability limits to
guessing.

Now we use the above-described procedure, {modified to
Helstrom discrimination, to bound the working probability
$\wp(\checkmark)$ of an immaculate amplification device.} The device
takes $|\alpha\rangle$ to $|g\alpha\rangle$ and $|\beta\rangle$ to
$|g\beta\rangle$.  Amplification increases the distinguishability of
the states so that the probability of successful identification of
the state is
\begin{align}
P^{\rm A}(\checkmark)
=\frac{1}{2} \! \left ( 1 + \sqrt{1- e^{-g^2|\alpha-\beta|^2}} \right),
\end{align}
where the superscript ``A'' reminds us this is {\em after\/}
amplification.  The overall probability to identify the input state
correctly after amplification is
\begin{align}
P_{\rm Hel}^{\rm A}(\checkmark)
&= \frac12 [1-\wp(\checkmark)]
+ \wp(\checkmark) P^{\rm A}(\checkmark)\nonumber\\
&=\frac12 \left(1+\wp(\checkmark)\sqrt{1- e^{-g^2|\alpha-\beta|^2}}\right)\,.
\end{align}
Since the probability for successful discrimination cannot increase,
we must have $P_{\rm Hel}^{\rm A}(\checkmark) \le P_{\rm Hel}^{\rm
B}(\checkmark)$, which gives an upper bound on the amplifier's
success probability,
\begin{align}
\wp(\checkmark)
\le  \sqrt{\frac{1- e^{-|\alpha-\beta|^2}}
{\vphantom{\Big)}1-e^{-g^2|\alpha-\beta|^2}} }\,.
\label{eq:Helpbound}
\end{align}
This bound, which holds for any pair of states, has its minimum value
when the two coherent states become very close to each other, i.e.,
$|\alpha - \beta|\rightarrow 0$; in this case the bound on the
working probability becomes
\begin{align}
\wp_{\rm Hel} \le \frac{1}{g}\,.
\label{eq:2Hel_bound}
\end{align}

For constructing models of immaculate amplifiers, Helstrom-type
discrimination has the problem that it sometimes misidentifies the
input state.  Such misidentification inevitably leads to noise in the
amplifier output, which cannot be part of a model of an exact
immaculate amplifier.

\subsection{USD bounds}

\subsubsection{Two coherent states}

Unambiguous state discrimination does discriminate states without
error, but this providence requires a sacrifice, namely, the
no-decision measurement result.   For two input states,
$\ket{\alpha}$ and $\ket{\beta}$, the probability of successfully
identifying them is~\cite{WisMilBook}
\begin{align}
P_{\rm USD}^{\rm B}(\checkmark)
= 1 - |\ip{\beta}{\! \alpha}|^2
= 1 - e^{-|\alpha-\beta|^2}\,.
\end{align}
In this expression, as in the Helstrom case, it is apparent that as
the separation, $|\alpha-\beta|$, between the two states grows, the
probability of discrimination approaches unity.  When the states get
close together, $|\alpha-\beta|\rightarrow 0$, the probability of
successful discrimination goes to zero.

After amplification we have a discrimination probability,
\begin{align}
P^{\rm A}(\checkmark)
&= 1 - |\ip{g\beta}{\! g \alpha}|^2
&= 1 - e^{-g^2|\alpha-\beta|^2}\,,
\end{align}
and an overall probability for successfully identifying the input
state,
\begin{align}
P_{\rm USD}^{\rm A}(\checkmark) &=\wp(\checkmark)\,P^{\rm A}(\checkmark) .
\end{align}
Since amplification cannot increase the distinguishability of the
states, we have $P_{\rm USD}^{\rm A}(\checkmark) \le P_{\rm USD}^{\rm
B}(\checkmark)$ and thus an upper bound on the working probability,
\begin{align}
\wp(\checkmark)\le\frac{P^B_{\rm USD}(\checkmark)}{P^A(\checkmark)}=
\frac{1- e^{-|\alpha-\beta|^2}}{\vphantom{\Big(}1- e^{-g^2|\alpha-\beta|^2}}\,,
\end{align}
as pointed out in Ref.~\cite{XiaRalLun10}.  Being the square of the
Helstrom bound~(\ref{eq:Helpbound}), this is always the tighter
bound.  The minimum of the bound is found in the limit that the
coherent states become very close to each other, i.e.,
$|\alpha-\beta|\rightarrow 0$, in which case the bound becomes
\begin{align}
\wp_{\rm USD} \le \frac{1}{g^2},\label{eq:2USD_bound}
\end{align}
The allowed working probability is a factor of $1/g$ smaller than the
Helstrom bound~(\ref{eq:2Hel_bound}).  This USD bound is the same as
the bound~(\ref{eq:Pmu}), which was derived by considering how to
distinguish neighboring coherent states using quadrature
measurements; the two bounds are the same because both are based on
discriminating neighboring coherent states.

\subsubsection{$M$ coherent states on a circle}

The USD bound~(\ref{eq:2USD_bound}) is not at all a tight bound on
the working probability for a probabilistic immaculate amplifier.  We
can get much tighter bounds by applying USD to more than two input
states.  Indeed, we work toward a phase-insensitive amplifier, which
must act symmetrically on {\em all\/} input coherent states with the
same $|\alpha|$.  Thus what we do is to consider a set of $M$
coherent states, $\ket{\alpha_j}=\ket{\bar\alpha e^{i\phi_j}}$, all
located on a circle of radius $\bar\alpha$ with phases
\begin{equation}
\phi_j=\frac{2\pi j}{M}\,,
\quad j=0,1,2,\ldots, M-1,
\end{equation}
distributed uniformly around the circle.  To avoid clutter in what
follows, we use, as here, $\bar\alpha=|\alpha|$.  To apply USD to the
states $|\alpha_j\rangle$, they must be linearly independent. This
property was shown in Ref.~\cite{ChaBenHel89}, and it emerges
naturally as part of the USD construction.  In contrast, the
continuum of states on the circle are complete, spanning the entire
Hilbert space, but are not linearly independent; we review these
facts in Appendix~\ref{sec:independence_day}.

Chefles and Barnett~\cite{CheBar98} solved the USD problem for sets
of linearly independent symmetric states (see also~\cite{Che98}). For
the case of coherent states on a circle, the unitary operator that
rotates between states is the phase-plane rotation by angle $2\pi/M$,
i.e., $U=e^{i2\pi a^\dagger a/M}$.  Restricted to the subspace
spanned by the set of input coherent states, $U$ has the
eigendecomposition
\begin{align}
U
=\sum_{r = 0}^{M-1} e^{i\phi_r}\ket{\gamma_r}\bra{\gamma_r}\,,
\end{align}
where the (orthonormal) eigenstates are given by
\begin{align}
c_r\ket{\gamma_r}
=\frac{1}{M}\sum_{j=0}^{M-1}e^{-i2\pi rj/M}\ket{\alpha_j}\,.
\end{align}
Here $c_r$, chosen to be real, is the magnitude of the vector on the
right:
\begin{align}
c_r^2
&= \frac{1}{M}\sum_{j=0}^{M-1}e^{-i2\pi rj/M}\ip{\alpha_0}{\alpha_j} \nonumber \\
&= \frac{1}{M}\sum_{j=0}^{M-1}e^{-i r\phi_j}\exp\bigl[\bar\alpha^2(e^{i\phi_j}-1)\bigr]\,.
\end{align}
It is useful to manipulate $c_r^2$ into a quite different form and
also to write it in terms of
\begin{align}
q_r=Mc_r^2&=e^{-\bar\alpha^2}\biggl.\frac{d^{M-r}}{dx^{M-r}}
\sum_{j=0}^{M-1}\exp\!\left(xe^{i\phi_j}\right)\biggr|_{x=\bar\alpha^2}\nonumber\\
&=Me^{-\bar\alpha^2}\sum_{k=0}^{\infty}\frac{\bar\alpha^{2(kM+r)}}{(kM+r)!}\,.
\label{eq:Pr}
\end{align}
That the states~$|\gamma_r\rangle$ are orthonormal establishes that
they and the original coherent states $|\alpha_j\rangle$ span an
$M$-dimensional subspace and thus that the $|\alpha_j\rangle$ are
linearly independent.

The vectors
\begin{align}
\ket{\alpha_j^{\perp}}
=\frac{1}{M}\sum_{r=0}^{M-1}\frac{1}{c_r}e^{i2\pi rj/M}\ket{\gamma_r}
\end{align}
are reciprocal (or dual) to the original coherent states in the sense
that $\langle\alpha_j^\perp|\alpha_k\rangle=\delta_{jk}$.  This
duality property is what is needed to construct the USD positive-operator-valued measure (\hbox{POVM}).
This POVM has $M$ POVM elements
$E_j=P(\checkmark)\ket{\alpha_j^\perp}\bra{\alpha_j^\perp}$,
$j=0,\ldots,M-1$, for the results that identify the input states,
where $P(\checkmark)$ is the success probability, and a single
failure POVM element, $E_{\rm fail}=I-E$, where
\begin{align}
E = \sum_jE_j=P(\checkmark)\sum_r \frac{1}{q_r}\ket{\gamma_r}\bra{\gamma_r}\,.
\end{align}

The largest eigenvalue of $E$ must be no larger than 1, which gives
an optimal success probability for discriminating among $M$ coherent
states symmetrically placed on a circle of radius
$\bar\alpha$~\cite{CheBar98}:
\begin{align}\label{eq:pusdmin}
P(\checkmark|\bar\alpha,M)=
\min_{\substack{r\in\{0,\ldots,M-1\}}}q_r\,,
\end{align}
This success probability has two important limits: (i)~many states on
the circle or, equivalently, small coherent-state amplitude, i.e.,
$M\gg\bar\alpha^2$, and (ii)~states sparse on the circle or,
equivalently, large coherent-state amplitude, i.e., $M\ll\bar\alpha$.
The reason for the difference in powers of $\bar\alpha$ in the two
limits {emerges} as we examine each limit in turn.

Notice first that the sums for $q_r/M$ in Eq.~(\ref{eq:Pr}) consist
of terms drawn with period $M$ from a Poisson distribution that has mean $\bar\alpha^2$, a distribution we denote
throughout by
$\mbox{Pr}[\,n\mid\bar\alpha^2\,]=e^{-\bar\alpha^2}\bar\alpha^{2n}/n!$.  
When the first term in the sum for $r=M-1$ lies beyond the maximum of
the Poisson distribution, as it does in the case of many states on
the circle, it takes only a moment's contemplation to realize that
the terms in the sum for $q_{M-1}$ are term by term smaller than the
corresponding terms in the sums for other values of $r$, provided
that the first term in $q_{M-1}$ is smaller than the first term in
$q_0$, i.e., $\bar\alpha^{2(M-1)}/(M-1)!<1$, which is certainly true
when $M\gg\bar\alpha^2$.  Thus, for many coherent states on the
circle, the minimum in Eq.~(\ref{eq:pusdmin}) is achieved by
$r=M-1$~\cite{CheBar98}, so
\begin{align}\label{eq:pandos}
P(\checkmark|\bar\alpha,M)=q_{M-1}
= M e^{-\bar\alpha^2}\sum_{k=0}^{\infty}\frac{\bar\alpha^{2(kM+M-1)}}{(kM+M-1)!}\,.
\end{align}
Moreover, the Chernoff bound for a Poisson random variable $n$ with
mean $\bar\alpha^2$~\cite{ChernoffPoisson}, applied to the terms in
the sum~(\ref{eq:pandos}) after the first,
\begin{align}
\sum_{k=1}^{\infty}\frac{\bar\alpha^{2(kM+M-1)}}{(kM+M-1)!}
&<e^{\bar\alpha^2}\mbox{Pr}[\,n\ge 2M-1\mid\bar\alpha^2\,]\nonumber\\
&\le\left(\frac{e\bar\alpha^2}{2M-1}\right)^{\!2M-1}\,,
\label{eq:Chernoff}
\end{align}
shows that, in the limit $M\gg\bar\alpha^2$, we need to keep only the
first term, $k=0$, of the sum~(\ref{eq:pandos}).  The result is a
simple expression for USD success probability in the case of many
coherent states on a circle (small coherent-state amplitude):
\begin{align}\label{eq:limitpandos}
P(\checkmark|\bar\alpha,M)=\frac{M e^{-\bar\alpha^2}\bar\alpha^{2(M-1)}}{(M-1)!}\,,
\quad M\gg\bar\alpha^2\,.
\end{align}

Now consider the case of sparse coherent states on the circle.   For
fixed $M$, as $\bar\alpha\rightarrow \infty$, Chefles and
Barnett~\cite{CheBar98} showed that all of the $q_r$ limit to 1, so
\begin{align}
P(\checkmark|\bar\alpha,M)= 1\,.
\end{align}
Since, for fixed $M$, the input states limit to being orthogonal as
$\bar\alpha\rightarrow\infty$, this simply means that orthogonal
states can be discriminated with unity probability of success.  More
useful than the limit, however, is the correction to the limit.

To find this correction, we begin by noting that since $\bar\alpha\gg
M\ge2$, we can approximate the Poisson distribution in \erf{eq:Pr} as
a Gaussian of the same mean and variance and extend the sum on $k$ to
$-\infty$ on the grounds that the Gaussian is negligible for these
additional terms:
\begin{align}
q_r=
\frac{M}{\sqrt{2\pi}\bar\alpha}\sum_{k=-\infty}^\infty
\exp\!\left(-\frac{(kM+r-\bar\alpha^2)^2}{2\bar\alpha^2}\right)\,.
\end{align}
By introducing $\delta$-functions, we can write this in the form
\begin{align}
q_r
&=\frac{\,M}{\sqrt{2\pi}\bar\alpha}\sum_{k=-\infty}^{\infty}
\int_{-\infty}^\infty dx\,e^{-(x-\bar\alpha^2)^2/2\bar\alpha^2}
\delta(x-kM-r)\nonumber\\
&=\frac{1}{\sqrt{2\pi}}
\int_{-\infty}^\infty du\,e^{-u^2/2}
\sum_{k=-\infty}^{\infty}\delta\!\left(k-\frac{\bar\alpha}{m}+\frac{s-u}{m}\right)\,,
\label{eq:gaus_comb}
\end{align}
where $x$ is a continuous version of $kM+r$, and where in the second
expression we introduce the integration variable
$u=x/\bar\alpha-\bar\alpha$ and rescaled variables
$m=M/\bar\alpha\ll1$ and $s=r/\bar\alpha\ll1$.  Now we write
$\bar\alpha/m=[\bar\alpha/m]+\aleph$, where $[z]$ denotes the nearest
integer to $z$ and, hence, $-\frac12\le\aleph<\frac12$ (half-integers
are rounded up), redefine the dummy summing variable to be
$k-[\bar\alpha/m]$, and use 
\begin{align}
\sum_{k=-\infty}^\infty\delta(k-v)
=\sum_{j=-\infty}^\infty e^{-i2\pi jv}
\end{align}
to put \erf{eq:gaus_comb} in the form
\begin{align}
q_r&=\frac{1}{\sqrt{2\pi}}\sum_{j=-\infty}^{\infty}e^{i2\pi j(s/m-\aleph)}
\int_{-\infty}^\infty du\,e^{-u^2/2}e^{-i2\pi ju/m}\nonumber\\
&=1+2\sum_{j=1}^{\infty}\cos\!\left[2\pi j\biggl(\frac{s}{m}-\aleph\biggr)\right]e^{-2\pi^2j^2/m^2}\nonumber\\
&=\theta_3\!\left[\pi\Big(\frac{s}{m}-\aleph\Big);e^{-2\pi^2/m^2}\right]\,.
\label{eq:theZhang_crazy}
\end{align}
Here $\theta_3$ denotes a Jacobi theta function~\cite{AandS}.

When $m \ll 1$, we only need to keep the $j=1$ term in the sum to get
the dominant correction to unity in $q_r$.  To minimize $q_r$, we
choose $r/M-\aleph=s/m-\aleph$ as close to $\frac12$ as possible,
consistent with letting $r$ be an integer.  Thus we choose
$r=[M(\aleph+\frac12)]$, which gives
\begin{align}
\cos\!&\left[2\pi\biggl(\frac{s}{m}-\aleph\biggr)\right]\nonumber\\
&=-1+\left(\mbox{irrelevant errors of size}\,\alt\frac{\pi^2}{2M^2}\right)\,.
\end{align}
{Keeping more terms in the sum and then minimizing could
provide a better approximation, but the lowest-order, $j=1$
correction already provides a good approximation for a reasonably
dense set of coherent states so the following analysis is restricted
to it.}

\begin{figure}[htbp]
\includegraphics[width=0.47\textwidth]{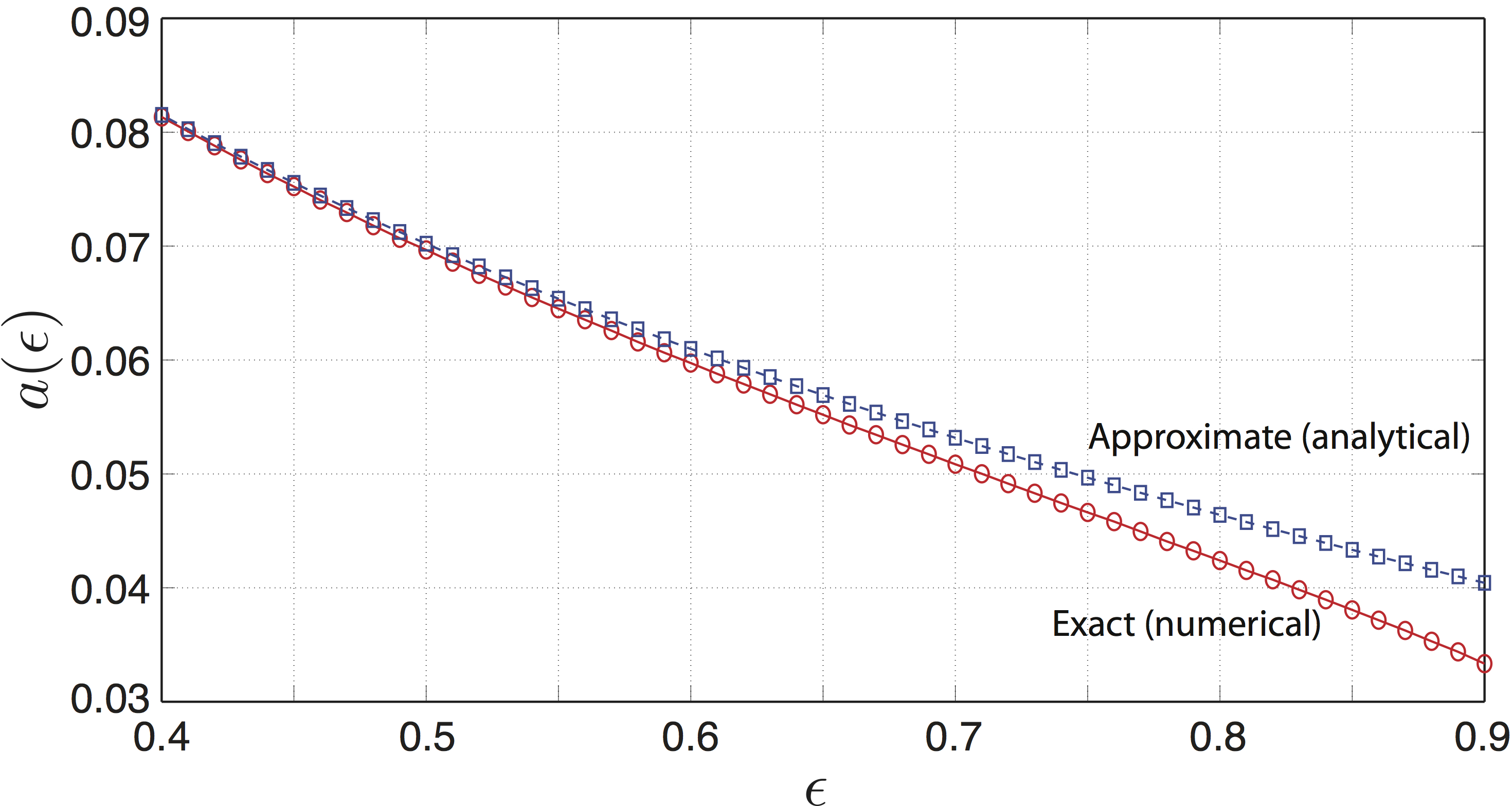}
\caption{(Color online) Dependence of the ratio
$\bar\alpha^2/M^2=a(\epsilon)$ on the deviation $\epsilon$ of the
success probability $P(\checkmark|\bar\alpha,M)$ from unity: numerical
results are plotted as (red) circles; analytic approximation of
Eq.~(\ref{eq:aepsilon}) as (blue) squares.  The analytic
approximation works quite well for $\epsilon\in [0, 0.5]$, but breaks
down progressively beyond $\epsilon=0.5$. }\label{fig3}
\end{figure}

The resulting success probability in the case of sparse coherent
states (large coherent-state amplitudes) is
\begin{align}\label{eq:theZhang}
P(\checkmark|\bar\alpha,M)=1-\epsilon\simeq1-2e^{-2\pi^2\bar\alpha^2/M^2}\,,
\quad
M\ll\bar\alpha\,.
\end{align}
The key result here is that in this limit the success probability
only depends on the ratio $\bar\alpha/M$.  Indeed, using this
expression, we can turn the question around and determine the ratio
that gives a deviation $\epsilon$:
\begin{align}
\frac{\bar\alpha^2}{M^2}
\equiv a(\epsilon)\simeq-\frac{\ln(\epsilon/2)}{2\pi^2}
=-0.05066\ln\epsilon+0.0351\,.
\label{eq:aepsilon}
\end{align}
For example, to achieve $P(\checkmark|\bar\alpha,M)=0.9$ for any $M$,
one chooses $\bar\alpha^2\simeq0.15M^2$.  The
dependence~(\ref{eq:aepsilon}) has been tested numerically over the
ranges $\epsilon\in[0.5,10^{-5}]$ and $M\in[2,40]$; the numerics give
\begin{align}
\label{eq:theZhang_a_epsilon}
a(\epsilon)= -0.0508\ln\epsilon+0.035\,,
\end{align}
in good agreement with the analytic approximation. Figure~\ref{fig3}
compares the numerics with the analytic approximation; the analytic
approximation works quite well for $\epsilon\le0.5$.

\begin{figure}[htbp]
\includegraphics[width=0.47\textwidth]{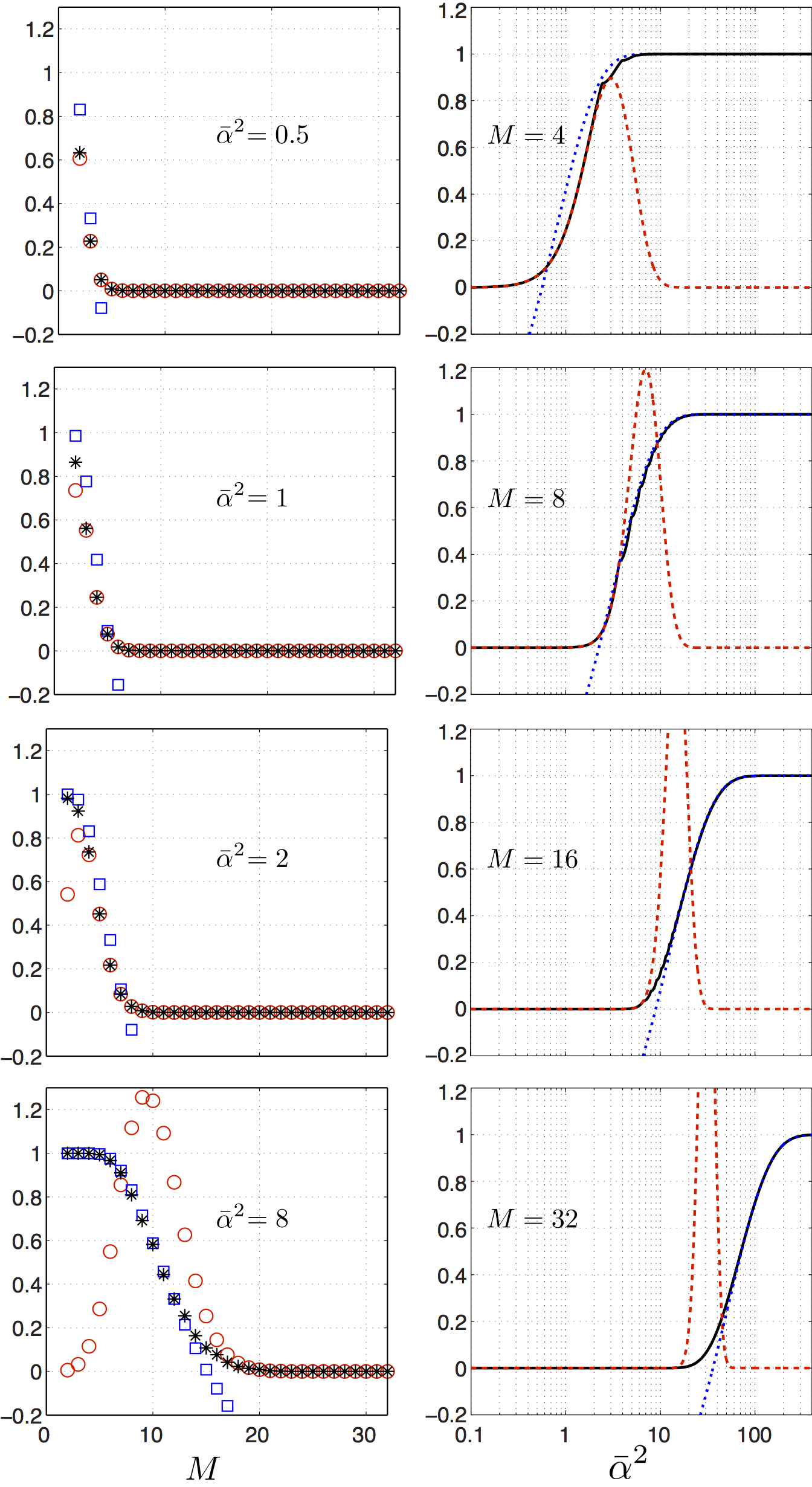}
\caption{(Color online) Success probability
$P(\checkmark|\bar\alpha,M)=P^{\rm B}(\checkmark)$. Left column: as a function
of $M$ with fixed $\bar\alpha^2$; (black) asterisks are the exact,
numerically determined success probability~(\ref{eq:pusdmin}); (red)
circles give the approximate result~(\ref{eq:limitpandos}) for many
coherent states (small coherent-state amplitude); (blue) squares give
the approximate result~(\ref{eq:theZhang}) for sparse coherent states
(large coherent-state amplitude).  Right column: as a function of
$\bar\alpha^2$ with fixed $M$; (black) solid line is the exact
result; (red) dashed line, many coherent states; (blue) dotted line,
sparse coherent states.}
\label{fig4}
\end{figure}

Figure~\ref{fig4} plots the success probability for USD of coherent
states on a circle, comparing the exact, numerically determined
result with the approximations that apply for many coherent states
and sparse coherent states.  The two approximations work better than
we have any right to expect: the plots and a consideration of the
next term in the sum~(\ref{eq:pandos}) suggest that the
many-coherent-states approximation~(\ref{eq:limitpandos}) works well
for $M\agt2\bar\alpha^2$; provided $\bar\alpha$ is somewhat bigger
than 1, the sparse-coherent-states approximation~(\ref{eq:theZhang})
works well for $M\alt4\bar\alpha$.  The two approximations overlap
when $\bar\alpha\agt1$ and $M$ are both small, but because of the
different powers of $\bar\alpha$ in the two approximations, generally
there is a gap between the two that must be filled in with numerics.

These results in hand, we can apply them, first, to obtain bounds on
the success probability of immaculate amplifiers and, second, to
constructing a model of an immaculate amplifier based on {USD}. For
the first task, we use the same notation as previously for before and
after probabilities of USD; the USD bound on the success probability
of an immaculate amplifier that works on the $M$ input coherent
states is
\begin{align}
\wp(\checkmark|\bar\alpha,M)
\le\frac{P^B_{\rm USD}(\checkmark)}{P^A(\checkmark)}
=\frac{P(\checkmark|\bar\alpha,M)}{P(\checkmark|g\bar\alpha,M)}\,.
\label{eq:pUSDsuccess}
\end{align}
The important cases of this bound require only our approximate
results for the USD success probabilities.

A first such case is when the input coherent states are sparse and,
hence, so are the amplified output states.  In this case, the
numerator and the denominator in the bound~(\ref{eq:pUSDsuccess}) are
both close to one, and the bound on success probability is also close
to one, reflecting the fact that one can discriminate and amplify
such nearly orthogonal states.

More interesting is the case of many input coherent states,
$M\gg\bar\alpha^2$.  If the gain is large enough that the amplified
states are sparse, i.e., $M\ll g\bar\alpha$---this requires that
$g\gg\bar\alpha$---the bound~(\ref{eq:pUSDsuccess}) reduces to
\begin{align}
\wp(\checkmark|\bar\alpha,M)\le P^B_{\rm USD}(\checkmark)
=\frac{M e^{-\bar\alpha^2}\bar\alpha^{2(M-1)}}{(M-1)!}
\label{eq:wpmanysparse}
\,.
\end{align}
This bound, which is plotted in Fig.~\ref{fig4} as (red) circles in
the left column and a (red) dashed line in the right column, can be
regarded as the $g\rightarrow\infty$ bound on an immaculate amplifier
that works on a fixed number $M\gg\bar\alpha^2$ of input states.

Most interesting is the case in which $M$ is large enough that both
the input and amplified output can be treated in the
many-coherent-states limit, i.e., $M\gg g^2\bar\alpha^2$.  In this
case, the bound~(\ref{eq:pUSDsuccess}) becomes
$\wp(\checkmark|\bar\alpha,M)\le e^{(g^2-1)\bar\alpha^2}/g^{2(M-1)}$.
This case is the most interesting because we can let $M$ become
arbitrarily large and thus approach the limit in which the amplifier
acts phase-insensitively on the entire circle of coherent states.
Since $M-1\gg(g^2-1)\bar\alpha^2$, we have
$e^{(g^2-1)\bar\alpha^2}\ll e^{M-1}$ and thus
\begin{align}
\wp(\checkmark|\bar\alpha,M)\le\frac{e^{(g^2-1)\bar\alpha^2}}{g^{2(M-1)}}
\ll\left(\frac{\sqrt e}{g^2}\right)^{2(M-1)}\,.
\label{eq:wpmanymany}
\end{align}
This shows that the success probability of an exact immaculate
amplifier goes to zero in the phase-insensitive limit
$M\rightarrow\infty$, even when the amplifier is only required to
work on a single circle of input coherent states.

We can make a more precise statement for an immaculate amplifier that
amplifies exactly all the coherent states on $M$ spokes
spaced equally in angle and of length $\bar\alpha$. Such an
amplifier acts immaculately on $M$ coherent states on all circles
with radius $\le\bar\alpha$. The success probability is bounded by
the $\bar\alpha\rightarrow0$ limit of the
bound~(\ref{eq:wpmanymany}), where the assumptions underlying the
bound are satisfied for any $M\ge2$:
\begin{align}
\wp_{\rm USD} \le \frac{1}{g^{2(M-1)}}\,.
\label{usd_bound}
\end{align}
This is one of the two chief results of this section: {\em an
immaculate amplifier that works exactly on $M$ spokes within a
phase-space disk centered at the origin has a working probability
that decreases exponentially with $M$, with the base of the
exponential, $g^2$, given by the gain, and goes to zero in the
phase-insensitive limit $M\rightarrow\infty$}.

It is useful to pause here to relate these results to the discussion
at the end of \srf{sec:laprior}.  For the disk amplifier, the
measurement-based performance measure~(\ref{eq:up}), which uses
antinormal ordering to calculate the uncertainties, is $1/\wp_{\rm
USD} g^2\ge g^{2(M-2)}$; this is greater than the
uncertainty-principle lower bound of one, achieved by an ideal linear
amplifier, for $M>2$ and far worse than the bound as $M$ gets large.
(These same arguments hold for the bound on the root-probability--SNR
product, which is equivalent to the uncertainty bound.)  The related
probability-fidelity product is given by $\wp_{\rm
USD}\le1/g^{2(M-1)}$, since an exact immaculate amplifier has unit
output fidelity; this is worse than the probability-fidelity product
$1/g^2$ achieved by an ideal linear amplifier for $M\ge2$ and far
worse as $M$ gets large.

As we discussed in the introductory paragraphs of this section, we
can construct a USD-based model of an immaculate amplifier in which
the $M$ input coherent states are first discriminated and then the
identified input is amplified immaculately by any amount.  The
quantum operation for this model is
\begin{align}
\mfA(\rho)=&
\sum_{j=0}^{M-1}\wp(\checkmark|\bar\alpha,M)
\ket{g\alpha_j}\bra{\alpha_j^{\perp}}\rho\ket{\alpha_j^{\perp}}\bra{g\alpha_j}\,.
\label{eq:Drho}
\end{align}
This map can be applied to any input state, not just the $M$ coherent
states used to construct it, but applied to one of those special
input states, $|\alpha_j\rangle$, $\mfA$ outputs the amplified state
$|g\alpha_j\rangle$ with probability
\begin{align}
\wp(\checkmark|\bar\alpha,M)=P^B_{\rm USD}(\checkmark)=P(\checkmark|\bar\alpha,M)\,.
\label{eq:wpB}
\end{align}
This success probability is plotted in Fig.~\ref{fig4}.

When $M\gg\bar\alpha^2$, the success probability is given by
Eq.~(\ref{eq:limitpandos}),
\begin{align}
\wp(\checkmark|\bar\alpha,M)\simeq
\sqrt{\frac{M}{2\pi}}e^{-\bar\alpha^2}\left(\frac{e\bar\alpha^2}{M-1}\right)^{M-1}\,,
\end{align}
where here we apply Stirling's approximation to the factorial to make
clear that the success probability goes to zero in the
phase-insensitive limit~$M\rightarrow\infty$.

The case of sparse input states is where immaculate amplification
shines with the radiance its name evokes.  As the plots in
Fig.~\ref{fig4} show, the success probability for this case is
captured by the sparse-states approximation~(\ref{eq:theZhang}),
which is plotted in Fig.~\ref{fig4} as (blue) squares in the left
column and a (blue) dotted line in the right column.  The
approximation works well for success probabilities
$1-\epsilon\agt0.5$, which corresponds to $M\alt4\bar\alpha$.  To
achieve a success probability $1-\epsilon$ requires that
$\bar\alpha/M=\sqrt{a(\epsilon)}$ be chosen as in
Eq.~(\ref{eq:aepsilon}).  To get a feeling for what these results
mean, notice that a success probability of $1-\epsilon$ corresponds
to a distance between states, measured along the arc of the circle,
given by $2\pi\bar\alpha/M=2\pi\sqrt{a(\epsilon)}$; for example, a
success probability of $0.5$ corresponds to $\sqrt a\simeq 0.265$ and
a distance of about $1.67$.  These states might seem pretty crowded,
but the distance makes sense when compared with the
one-standard-deviation diameter of a coherent state, which is 1.
These input states are just beginning to overlap, but they are far
enough apart that they can be distinguished and amplified
immaculately half the time.

The lesson here is important: {\em USD-based devices can outperform
ideal linear amplifiers if they are both phase-sensitive and
amplitude-specific, amplifying immaculately only a relatively sparse
set of input coherent states on a particular input circle.}  This
realization leads to a set of interesting questions that we consider
briefly in the Conclusion as the basis for future work. The flip side
is that success probability goes to zero when an exact immaculate
device is required to work phase-insensitively on even a single input
circle.  This suggests that phase insensitivity is a key property,
which does not play well with exact immaculate amplification. In the
next section, we explore this further by considering probabilistic
immaculate amplifiers that are required to be phase-insensitive, but
unlike USD-based amplifiers, are not exact.

\section{Bounds on phase-insensitive, approximate, probabilistic immaculate
amplification}\label{sec:kraus}

In this section we canonize phase insensitivity as a
primary requirement for amplification.  This means that the
amplifier's operation must be invariant under phase-plane rotations.
We relax the requirement of unit fidelity with the target output
state, thus obtaining a model of an approximate immaculate amplifier.
We would like the amplifier to work with high fidelity for input
coherent states $|\alpha\rangle$ within a disk centered at the
origin, but we allow the fidelity with the target amplified state
$|g\alpha\rangle$ to fall off for inputs outside the disk of
interest. {There are two motivations for this assumption:
first, as we noted in \srf{sec:paprior}, an immaculate amplifier
cannot work over the entire phase plane; second, as was true for the
implementations reviewed in Sec.~\ref{sec:paprior}, such a cutoff is
a property of practical devices.}

We characterize the high-fidelity output region as a disk of radius
$\sqrt{N}$; the corresponding input disk thus has radius
$\sqrt{N}/g$.  After translating this description into the language
of amplifier maps and Kraus operators, we characterize the amplifier
in terms of the fidelity with the target state, $F(\bar\alpha)$, and
the probability that the amplifier works, $p(\checkmark|\bar\alpha)$,
both of which are functions of the input amplitude
$\bar\alpha=|\alpha|$.  We maximize the fidelity at each $\bar\alpha$
given a working probability at that $\bar\alpha$, after which we
maximize the working probability consistent with the amplifier's map
being trace decreasing.  We thus obtain an optimal immaculate
amplifier that is both approximate and probabilistic.

We note that a similar analysis has been performed by
Fiur\'a\v{s}ek~\cite{Fiur01a,Fiur04a} in the context of cloning and
arbitrary state transformations; we point out below similarities to
and differences from our analysis.

We describe the amplification process by a quantum operation, which
we write in terms of a canonical Kraus decomposition in which the
Kraus operators are orthogonal.  We assume that these Kraus operators
have the form $P_NK_j$, where $P_N$ is the projector onto the
subspace $S_N$ spanned by the first $N+1$ number states.  The
amplifier quantum operation is thus
\begin{equation}
\sA_N=\sum_j P_NK_j\odot K_j^\dagger P_N\,,
\label{eq:sAN}
\end{equation}
where the $\odot$, technically a tensor product, can be regarded as
designating the slot for the input to the quantum operation.  The
projector $P_N$ provides a sharp cutoff in the number basis, beyond
which the amplifier's output has no support; notice that we can let
the operators $K_j$ map outside $S_N$ without having any effect on
the quantum operation~(\ref{eq:sAN}).  Shortly we extend the Kraus
operators in a way that allows the outputs to have support outside
$S_N$; this extension smooths the rough edges in the amplifier
map~(\ref{eq:sAN}), and it provides marginal improvements in the
output fidelity.  Phase insensitivity is the requirement that $\sA_N$
commutes with phase-plane rotations; this implies, as we show in
Appendix~\ref{sec:phasepreserve}, that each Kraus operator has
nonzero number-basis matrix elements on only one diagonal strip, as
in Eq.~(\ref{eq:symmKraus}).  Additionally, the Kraus operators must
satisfy the trace-decreasing requirement,
\begin{align}
\sum_j K_j^\dagger P_N K_j\le I\,.
\end{align}

Suppose now that the input state to the amplifier is a coherent state
$|\alpha\rangle$. The probability of outcome $j$ is
\begin{equation}
p_j(\checkmark|\bar\alpha)=\langle\alpha|K_j^\dagger P_NK_j|\alpha\rangle\,,
\end{equation}
and the overall success probability is
\begin{equation}
p(\checkmark|\bar\alpha)=\sum_jp_j(\checkmark|\bar\alpha)
=\tr\bigl[\sA_N\bigl(|\alpha\rangle\langle\alpha|\bigr)\bigr]\,.
\end{equation}
The fidelity of the output with the target output state
$|g\alpha\rangle$ is
\begin{equation}
F(\bar\alpha)
=\frac{\bigl\langle g\alpha\bigl|\sA_N\bigl(|\alpha\rangle\langle\alpha|\bigr)\bigr|g\alpha\bigr\rangle}
{p(\checkmark|\alpha)}\,.
\end{equation}
Because of the rotational symmetry, these quantities depend only on
the magnitude $\bar\alpha=|\alpha|$.

The problem we solve is the following: fix a circle of coherent
states with amplitude $\bar\alpha$, and find the maximum fidelity
$F(\bar\alpha)$ on this circle for a fixed success probability
$q=p(\checkmark|\bar\alpha)$.  We do this first for a single Kraus
operator and later argue that a single Kraus operator is better than
more than one.  The optimization problem is thus to maximize
\begin{align}
F(\bar\alpha)=
\frac{|\langle g\alpha|P_NK|\alpha\rangle|^2}{p(\checkmark|\bar\alpha)}\,,
\end{align}
subject to the constraint
\begin{align}
q=p(\checkmark|\bar\alpha)=\langle\alpha|K^\dagger P_NK|\alpha\rangle\,.
\end{align}
We can, of course, rephrase this as maximizing $|\langle
g\alpha|P_NK|\alpha\rangle|^2$ subject to the constraint on working
probability.

Introducing a Lagrange multiplier $\mu$, we maximize
\begin{align}
|\langle g\alpha|P_NK|\alpha\rangle|^2
-\mu\bigl(\langle\alpha|K^\dagger P_NK|\alpha\rangle-q\bigr)\,.
\end{align}
Varying $K$ gives
\begin{align}
0=&\langle\alpha|\delta K^\dagger
\Bigl(P_N|g\alpha\rangle\langle g\alpha|P_NK|\alpha\rangle
-\mu P_NK|\alpha\rangle\Bigr)\nonumber\\
&+\mbox{(Hermitian conjugate)}\,,
\end{align}
so we conclude that
\begin{equation}
P_NK|\alpha\rangle=
P_N|g\alpha\rangle\frac{\langle g\alpha|P_NK|\alpha\rangle}{\mu}\,.
\end{equation}
The Lagrange multiplier is given by the probability for the first
$N+1$ photons in the target state $|g\alpha\rangle$,
\begin{align}
\mu&=\bra{g\alpha}P_N\ket{g\alpha}
=e^{-g^2|\alpha|^2}e_N(g^2|\alpha|^2)\,,
\end{align}
where we introduce a standard shorthand for the first $N+1$ terms in
the expansion of the exponential,
\begin{align}
e_N(x) \equiv \sum_{n=0}^{N} \frac{x^n}{n!}\,.
\end{align}
Without changing the Kraus operator $P_NK$, we can let $K$ map
outside the subspace $S_N$ in such a way that
\begin{equation}
K|\alpha\rangle=|g\alpha\rangle\frac{\langle g\alpha|P_NK|\alpha\rangle}{\mu}\,.
\end{equation}
Since
\begin{align}
g^{a^\dagger
a}|\alpha\rangle=e^{(g^2-1)|\alpha|^2/2}|g\alpha\rangle\,,
\end{align}
we can simplify this by letting $K=Lg^{a^\dagger a}$.  The result,
\begin{equation}
L|g\alpha\rangle=|g\alpha\rangle\frac{\langle g\alpha|P_NL|g\alpha\rangle}{\mu}\,,
\end{equation}
says that $|g\alpha\rangle$ is an eigenstate of $L$.  Since the
coherent states on a circle are a basis for the Hilbert space, this
determines $L$ to be a function of the annihilation operator $a$. The
rotational symmetry further requires that $L$ have number-state
matrix elements on only one diagonal strip, implying that $L=\lambda
a^k$, where $k$ is a nonnegative integer and $\lambda$ can be taken
to be real without loss of generality.

The possible optimal Kraus operators are
\begin{align}
K_k&=\lambda a^k g^{a^\dagger a}
=\lambda g^{a^\dagger a}(ga)^k\nonumber\\
&=\lambda\sum_{n=0}^\infty g^{n+k}\sqrt{\frac{(n+k)!}{n!}}|n\rangle\langle n+k|\,,\nonumber\\
&\qquad\qquad k=0,1,2,\ldots\,.
\label{eq:Kk1}
\end{align}
This operator has nonzero matrix elements only on the $k$th diagonal
strip above the main diagonal.  It is not surprising that this class
of operators emerges, because they do take $|\alpha\rangle$ to a
multiple of $|g\alpha\rangle$, just as we would like an immaculate
amplifier to do.  The success probability and fidelity become
\begin{align}
p(\checkmark|\bar\alpha)
&=\lambda^2g^{2k}e^{(g^2-1)\bar\alpha^2}\bar\alpha^{2k}
\langle g\alpha|P_N|g\alpha\rangle\,,\\
F(\bar\alpha)&=\bra{g\alpha}P_N\ket{g\alpha}
=e^{-g^2\bar\alpha^2}e_N\bigl(g^2\bar\alpha^2\bigr)=\mu\,.
\label{eq:Fopt1}
\end{align}

We can increase the success probability without changing the fidelity
by letting $\lambda^2$ increase, but there is a limit to this
increase set by the requirement that
\begin{align}
I\ge K_k^\dagger P_N K_k
=\lambda^2\sum_{n=0}^N g^{2(n+k)}\frac{(n+k)!}{n!}|n+k\rangle\langle n+k|\,.
\label{eq:KkPNKk}
\end{align}
Since the eigenvalues increase with $n$, the constraint is set by the
largest eigenvalue ($n=N$).  Choosing the largest possible value,
\begin{equation}
\lambda^2=\frac{N!}{(N+k)!}\frac{1}{g^{2(N+k)}}\,,
\end{equation}
maximizes the success probability.

The final results of these considerations are the Kraus operators
\begin{align}
K_k=\sqrt{\frac{N!}{(N+k)!}}\frac{a^k g^{a^\dagger a}}{g^{N+k}}
=\sqrt{\frac{N!}{(N+k)!}}\frac{g^{a^\dagger a}a^k}{g^N}
\label{eq:Kk2}
\end{align}
and the corresponding success probability and fidelity,
\begin{align}
\label{eq:pkopt1}
p(\checkmark|\bar\alpha)
&=\frac{N!}{(N+k)!}\frac{e^{-\bar\alpha^2}\bar\alpha^{2k}}{g^{2N}}e_N(g^2\bar\alpha^2)\,,\\
\label{eq:Fopt2}
F(\bar\alpha)&=e^{-g^2\bar\alpha^2}e_N(g^2\bar\alpha^2)\,.
\end{align}

Equation~(\ref{eq:Fopt2}) was derived by
Fiur\'a\v{s}ek~\cite{Fiur04a} (our amplitude gain $g$ is his
$\sqrt{M}$) by maximizing an average fidelity.  Fiur\'a\v{s}ek
considers a Gaussian distribution of input coherent states.  His
average fidelity includes, first, an average over the success
probability, normalized to an average success probability, averaged
over the input Gaussian, and, second, an average over the input
Gaussian. He does not quote the probability of success, and he only
finds the $k=0$ case.  He formulates the optimization problem as a
semidefinite program, whereas we use a simple Lagrange-multiplier
maximization.

It is useful to pause here to summarize properties of the fidelity
and the success probability.  The fidelity~(\ref{eq:Fopt2}) is the
probability of the first $N+1$ number states in the Poisson
distribution associated with the coherent state $|g\alpha\rangle$. As
we anticipated, this fidelity is close to 1 for $g\bar\alpha\ll\sqrt
N$, goes to zero for $g\bar\alpha\gg\sqrt N$, and transitions between
these two extremes around $g\bar\alpha\simeq\sqrt N$.  Indeed, we can
use the Chernoff bound for the probability in the tails of a Poisson
distribution with mean $g^2\bar\alpha^2$ to bound the fidelity in the
two extremes~\cite{ChernoffPoisson}, 
\begin{align}
g^2\bar\alpha^2\le N:\,\,\,
F(\bar\alpha)&=1-\mbox{Pr}[\,n\ge N+1\mid g^2\bar\alpha^2\,]\nonumber\\
&\ge1-e^{-g^2\bar\alpha^2}\left(\frac{eg^2\bar\alpha^2}{N+1}\right)^{N+1}\\
g^2\bar\alpha^2>N:\,\,\,
F(\bar\alpha)&=\mbox{Pr}[\,n\le N\mid g^2\bar\alpha^2\,]\nonumber\\
&\le\left(\frac{eg^2\bar\alpha^2e^{-g^2\bar\alpha^2/N}}{N}\right)^N\,.
\end{align}
The width of the transition region can be estimated by remembering
that the two-standard-deviation phase-plane radius of a coherent
state is $1$. As a consequence, the amplified output begins to
contact the number state cutoff at $N$ when $g\bar\alpha+1\simeq\sqrt
N$ and leaves the high-fidelity region entirely when
$g\bar\alpha-1\simeq\sqrt N$. Thus we expect the transition from
unity fidelity to zero fidelity to occur as $\bar\alpha$ varies from
$(\sqrt N-1)/g$ to $(\sqrt N+1)/g$.

The fidelity does not depend on $k$, but the success probability
does, so the value of $k$ that maximizes the success probability can
change as $\bar\alpha$ changes.  The amplifier map~(\ref{eq:sAN})
cannot depend, of course, on the input amplitude, so we must settle
on a value of $k$ and apply the resulting map to all input coherent
states.  We are most interested in the high-fidelity regime, where
the leading-order behavior of the success
probability~(\ref{eq:pkopt1}) is
\begin{align}\label{eq:pklimit}
p(\checkmark|\bar\alpha)
&=\frac{N!}{(N+k)!}
\frac{e^{-\bar\alpha^2}\bar\alpha^{2k}}{g^{2N}}\,,
\quad \bar\alpha\ll\sqrt N/g\,.
\end{align}
In this regime all values of $k$ have success probabilities that are
exponentially small in $N$, but $k=0$ is the best of a sad lot,
indicating that it is the best value of $k$.  Before investigating
the different values of $k$ in detail, however, we extend the Kraus
operator $P_NK_k$ so that it can map outside $S_N$ in a way that
increases the fidelity and success probability.

The extension we seek should preserve the phase insensitivity of
$P_NK_k$ and should not interfere with the output of $P_NK_k$ in the
subspace $S_N$.  A glance at Eq.~(\ref{eq:Kk1}) shows that the
extension must have the form $\Upsilon_k = P_NK_k
+\sum_{n=N+1}^{\infty}\upsilon_n\ket{n}\bra{n+k}$.  Now we impose the
condition
\begin{align}
I\ge\Upsilon_k^{\dagger}\Upsilon_k
=K_k^\dagger P_NK_k
+\sum_{n=N+1}^{\infty}|\upsilon_n|^2\op{n+k}{n+k}\,.
\end{align}
The term $K_k^\dagger P_NK_k$ already satisfies the inequality in the
subspace $S_{N+k}$ spanned by the first $N+k+1$ number states [see
Eq.~(\ref{eq:KkPNKk})], and we can maximize the amplifier's success
probability by saturating the inequality for the second term, i.e.,
by choosing $\upsilon_n=1$ for $n=N+1,N+2,\ldots\;$, with the result
that
\begin{align}
\Upsilon_k = P_NK_k
+\sum_{n=N+1}^{\infty}\ket{n}\bra{n+k}\,.
\label{eq:Upsilonk}
\end{align}
With this choice, notice that for $k=0$, the additional term in
$\Upsilon_k$ is simply the unit operator in the orthocomplement of
$S_N$.

The extension of the Kraus operator has essentially no impact on the
operation of the amplifier in the high-fidelity input region.  It
does increase the fidelity marginally in the transition region by
including in the output number-state components with $n>N$.  The
biggest effect is to increase dramatically the success probability in
the low-fidelity regime beyond $\bar\alpha\simeq\sqrt{N+k}$, but this
improvement is a pyrrhic victory: all it does is to allow the
amplifier to report that it worked on inputs where the output has
essentially the same fidelity with the target as the input does.

Using the extended Kraus operators to calculate the success
probability and the fidelity of the output with the target
$|g\alpha\rangle$ gives
\begin{widetext}
\begin{align}
\label{eq:pkopt2}
p_k(\checkmark|\bar\alpha)
&=\bra{\alpha}\Upsilon_k^{\dagger}\Upsilon_k\ket{\alpha}
=e^{-\bar\alpha^2}\bar\alpha^{2k}
\!\left(
\frac{N!}{(N+k)!}\frac{1}{g^{2N}}e_N(g^2\bar\alpha^2)
+\sum_{n=N+1}^\infty\frac{\bar\alpha^{2n}}{(n+k)!}
\right)\,,\\
F_k(\bar\alpha)&=\frac{|\bra{g\alpha}\Upsilon_k\ket{\alpha}|^2}{p_k(\checkmark|\alpha)}
=\frac{e^{-g^2\bar\alpha^2}}{p_k(\checkmark|\bar\alpha)/e^{-\bar\alpha^2}\bar\alpha^{2k}}
\left(\sqrt{\frac{N!}{(N+k)!}}\frac{1}{g^N}e_N(g^2\bar\alpha^2)
+\sum_{n=N+1}^{\infty}\frac{g^n\bar\alpha^{2n}}{\sqrt{n!(n+k)!}}\right)^2\,.
\label{eq:Fkopt}
\end{align}
\end{widetext}
With the extended Kraus operators, both the success probability and
the fidelity depend on $k$.  In the high-fidelity regime,
$\bar\alpha\ll\sqrt N/g$, the extension terms have little impact: the
fidelity limits to unity, and the success probability has the form
given in Eq.~(\ref{eq:pklimit}), which decreases exponentially with
$N$.  For $\bar\alpha\gg\sqrt N/g$, the fidelity goes to zero much as
it did before.  The success probability, however, has a new
transition that occurs at $\bar\alpha^2\simeq N+k$: for
$\bar\alpha^2\gg N+k$, only the extension term matters, so the
success probability becomes nearly the entire probability under a
Poisson distribution with mean $\bar\alpha^2$, i.e.,
$p_k(\checkmark|\bar\alpha)=\mbox{Pr}[\,n\ge
N+k+1\mid\bar\alpha^2\,]$, and this limits to unity as
$\bar\alpha^2\rightarrow\infty$.

\begin{figure}[htbp]
\includegraphics[width=0.47\textwidth]{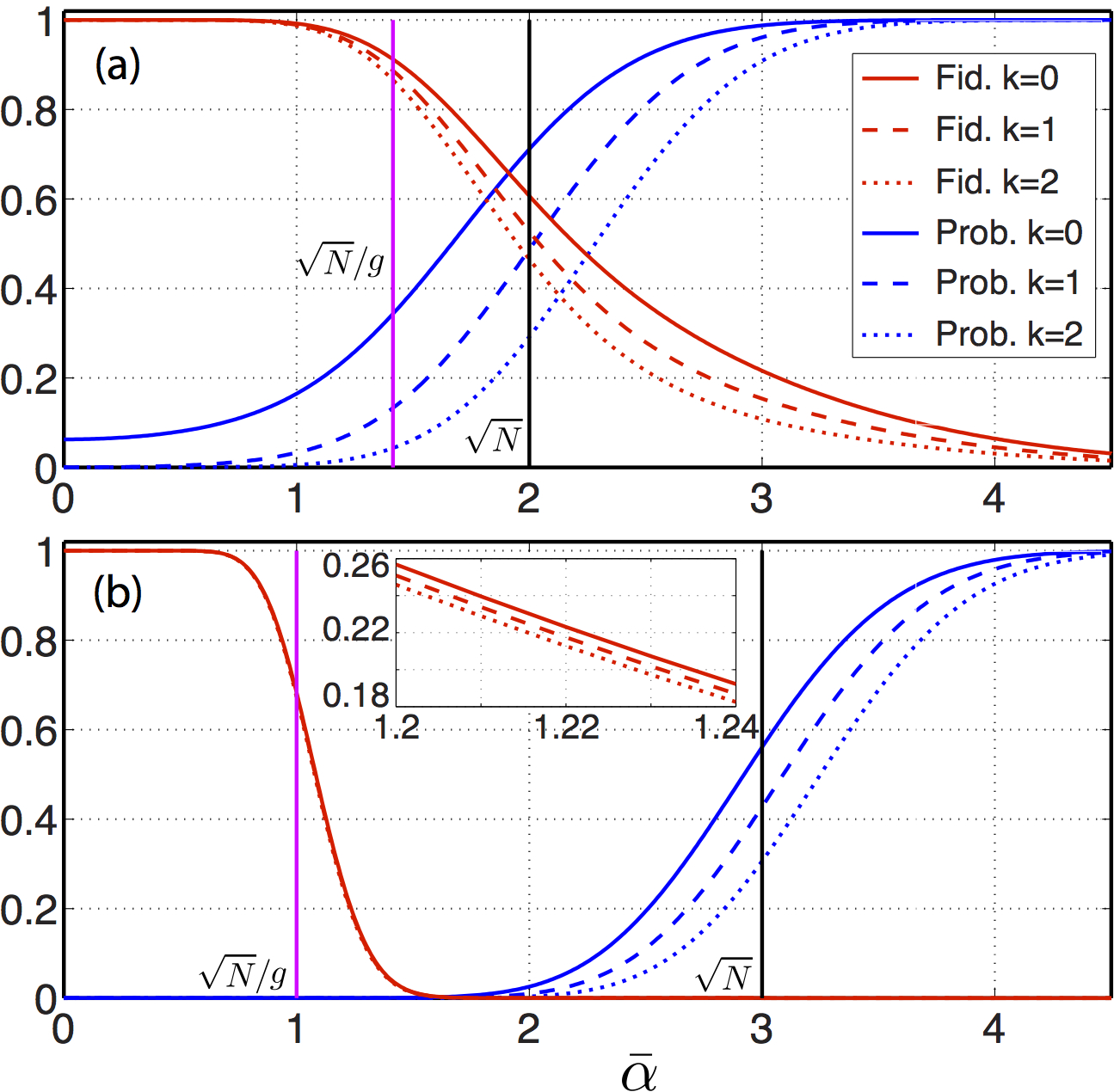}
\caption{(Color online) Fidelity $F_k(\bar\alpha)$ of
Eq.~(\ref{eq:Fkopt}) (descending curves) and success
probability~$p_k(\checkmark|\bar\alpha)$ of Eq.~(\ref{eq:pkopt2}) (ascending
curves) plotted as functions of input amplitude $\bar\alpha$ for
different extended Kraus operators $\Upsilon_k$ with $k=0$ (solid
lines), 1 (dashed lines), and 2 (dotted lines): (a)~ $g=\sqrt{2}$,
$N=4$; (b)~$g=3$, $N=9$.  The inset in~(b) illustrates the small
differences in fidelity, undetectable in the main plot, among the
three values of $k$.}\label{fig5}
\end{figure}

To gain insight into the success probability~(\ref{eq:pkopt2}) and
output fidelity~(\ref{eq:Fkopt}), we plot them in \frf{fig5} as a
function of the input amplitude $\bar\alpha$ for $k=0$, 1, and 2. In
\frf{fig5}(a), we take an amplitude gain $g=\sqrt2$ and $N=4$, both
of which are too small to see some of the characteristic features we
have discussed. The three fidelity curves are approximately unity
until $\bar\alpha\sim\sqrt{N}/g$. After this point the fidelity
decreases to zero.  Conversely, the three success-probability curves
start close to zero and rise to unity after $|\alpha|\sim\sqrt{N}/g$.
Figure~\ref{fig5}(b) plots the same curves for $g=3$ and $N=9$,
values big enough to see the characteristic features of the two
quantities.  In particular, it is apparent that the fidelity
transitions from unity fidelity to zero fidelity around
$\bar\alpha\simeq\sqrt{N}/g=1$, with the transition occurring between
$(\sqrt N-1)/g=2/3$ and $(\sqrt N+1)/g=4/3$, as anticipated.  For all
three values of $k$, the success probability in part~(c) rises from
its initial small value to unity, with the rise occurring around the
second transition at $\bar\alpha\simeq\sqrt{N}$.

It turns out that the success probability and fidelity for any value
of $k$ are bounded in the following way:
\begin{align}
\label{eq:pkbounds}
0\le p_k(\checkmark|\bar\alpha)&\le p_0(\checkmark|\bar\alpha)\,,\\
\label{eq:Fkbounds}
F_k(\bar\alpha)&\le F_0(\bar\alpha)\,.
\end{align}
These bounds are illustrated by the examples plotted in
Fig.~\ref{fig5}, and we have proven them analytically.  The proof,
which is tedious, is contained in Appendix~\ref{sec:optimal}. The
bounds confirm that the best value of $k$ is $k=0$.  We also show in
Appendix~\ref{sec:optimal} that
\begin{equation}
F_0(\bar\alpha)\ge\langle g\alpha|P_N|g\alpha\rangle\,,
\label{eq:F0ineq}
\end{equation}
which indicates that the $k=0$ extension increases the fidelity over
that of the restricted Kraus operators.

If the amplifier quantum operation has Kraus operators other than
$\Upsilon_0$, our analysis shows that these other Kraus operators
necessarily reduce the fidelity and the success probability.  This
justifies our earlier assumption of a single Kraus operator.  The
best Kraus operator is $\Upsilon_0$, and this gives an amplifier
quantum operation $\sA_N=\Upsilon_0\odot\Upsilon_0^\dagger$.

\begin{figure}[htbp]
\includegraphics[width=0.49\textwidth]{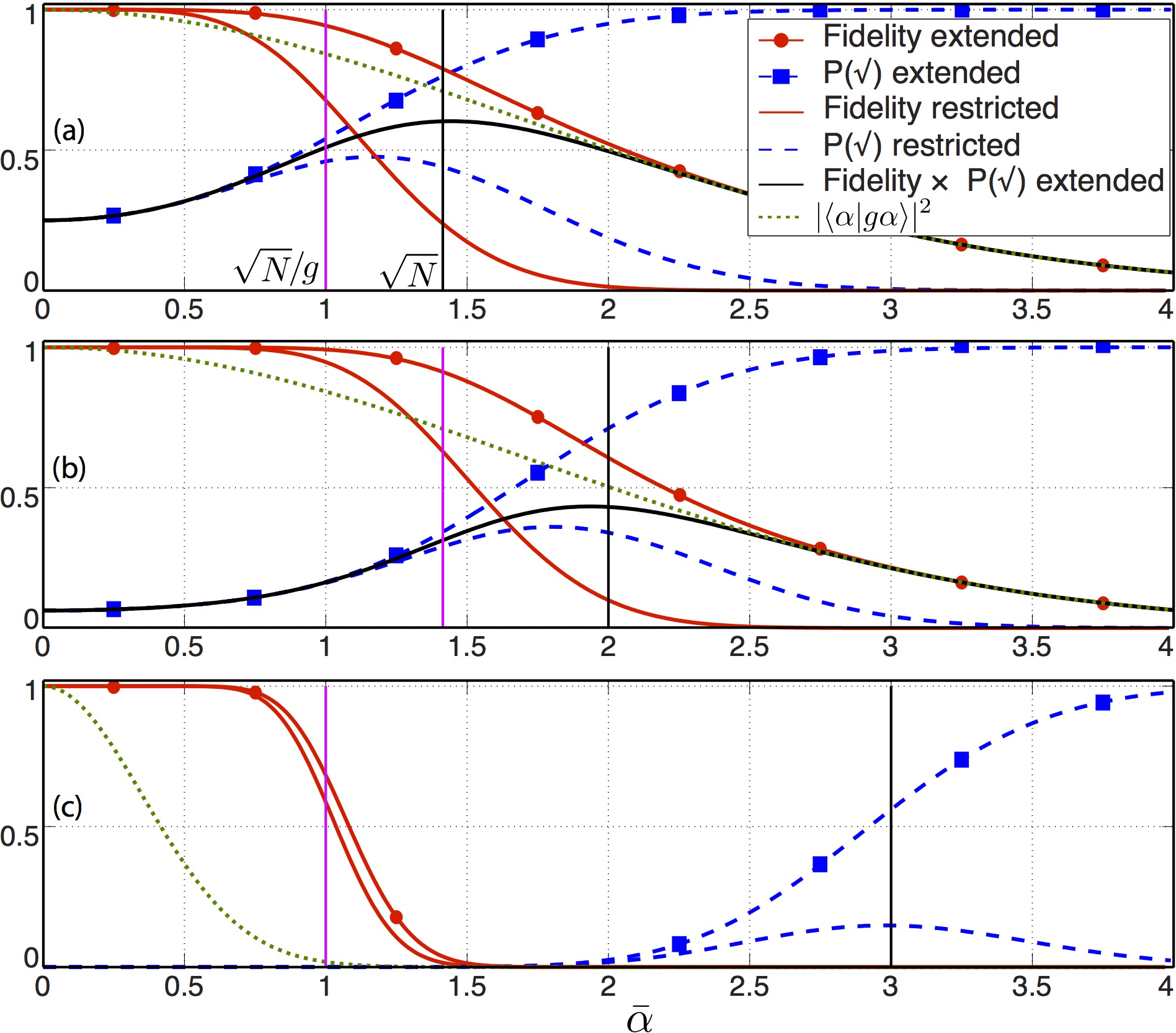}
\caption{(Color online) Fidelity $F_0(\bar\alpha)$ using the extended
Kraus operator $\Upsilon_0$ [Eq.~(\ref{eq:Fkopt})] (solid line with
filled circles); corresponding success
probability~$p_0(\checkmark|\bar\alpha)$ [Eq.~(\ref{eq:pkopt2})] (dashed
line with filled squares); fidelity $F(\bar\alpha)$ using the
restricted Kraus operator $P_NK_0$ [Eq.~(\ref{eq:Fopt2})] (solid
descending line); corresponding restricted success
probability~$p(\checkmark|\bar\alpha)$ [Eq.~\ref{eq:pkopt1}) with $k=0$]
(dashed line); probability-fidelity product
$p_0(\checkmark|\bar\alpha)F_0(\bar\alpha)$ (solid humped line); and
overlap $|\langle\alpha|g\alpha\rangle|^2$ (dotted line), all plotted
as functions of input amplitude~$\bar\alpha$: (a)~$g=\sqrt{2}$,
$N=2$; (b)~$g=\sqrt{2}$, $N=4$; (c)~$g=3$ $N=9$.}\label{fig6}
\end{figure}

The three plots in \frf{fig6}, all for $k=0$, have different values
of $g$ and $N$, but roughly the same high-fidelity input region: the
ratio $\sqrt{N}/g = 1$ in parts~(a) and~(c), whereas in~(b) it is
$\sqrt2$.  The plots include the fidelity and success probability
coming from the extended Kraus operator $\Upsilon_0$ and, for
comparison, the fidelity and success probability coming from the
restricted Kraus operator $P_NK_0$.  Parts~(a) and~(b) are
interesting because they have gains typical of that achieved in
experiments, but the transitions are not very sharp, $g$ and $N$
being too small to see the characteristic features of the plotted
quantities. In part~(c), where $g=3$ and $N=9$, the characteristic
features emerge: the extended Kraus operator provides a small
increase in fidelity through the transition region; the success
probability using $\Upsilon_0$ ascends to 1 beyond
$\bar\alpha\simeq\sqrt N$, instead of falling back to nearly zero as happens with the success probability that comes from
$P_NK_0$. These plots illustrate the superior qualities of the
extended Kraus operator $\Upsilon_0$; we do not consider the
restricted Kraus operators again.

Figure~\ref{fig6} plots two other quantities: the
probability-fidelity product,
$p_0(\checkmark|\bar\alpha)F_0(\bar\alpha)=|\langle
g\alpha|\Upsilon_0|\alpha\rangle|^2$, for our phase-insensitive
immaculate amplifier and the overlap
$|\langle\alpha|g\alpha\rangle|^2=e^{-(g-1)^2\bar\alpha^2}$.  The
latter can be regarded as the fidelity against the target state of a
device that does nothing, i.e., outputs the input.  Since nothing can
be done with unit probability, $|\langle\alpha|g\alpha\rangle|^2$ is
also the probability-fidelity product for a device that does nothing.
A minimal requirement for a useful amplifier is that it be better
than doing nothing.  The plots suggest that, as far as the
probability-fidelity product is concerned, the phase-insensitive
immaculate amplifier is never better than doing nothing---indeed,
$|\langle g\alpha|\Upsilon_0|\alpha\rangle|^2\le|\langle
g\alpha|\alpha\rangle|^2$ follows immediately from the fact that
$\Upsilon_0$ is diagonal in the number basis with positive
eigenvalues bounded above by 1---and approaches that standard only
for $\bar\alpha\agt\sqrt N$, where as we have already seen,
$\Upsilon_0$ becomes the identity map.  For comparison, the
probability-fidelity product for an ideal linear amplifier is $1/g^2$
[see Eq.~(\ref{eq:PFP})], which beats the do-nothing standard for
$\bar\alpha^2\ge\ln g^2/(g-1)^2$.

\begin{figure}[htbp]
\includegraphics[width=0.47\textwidth]{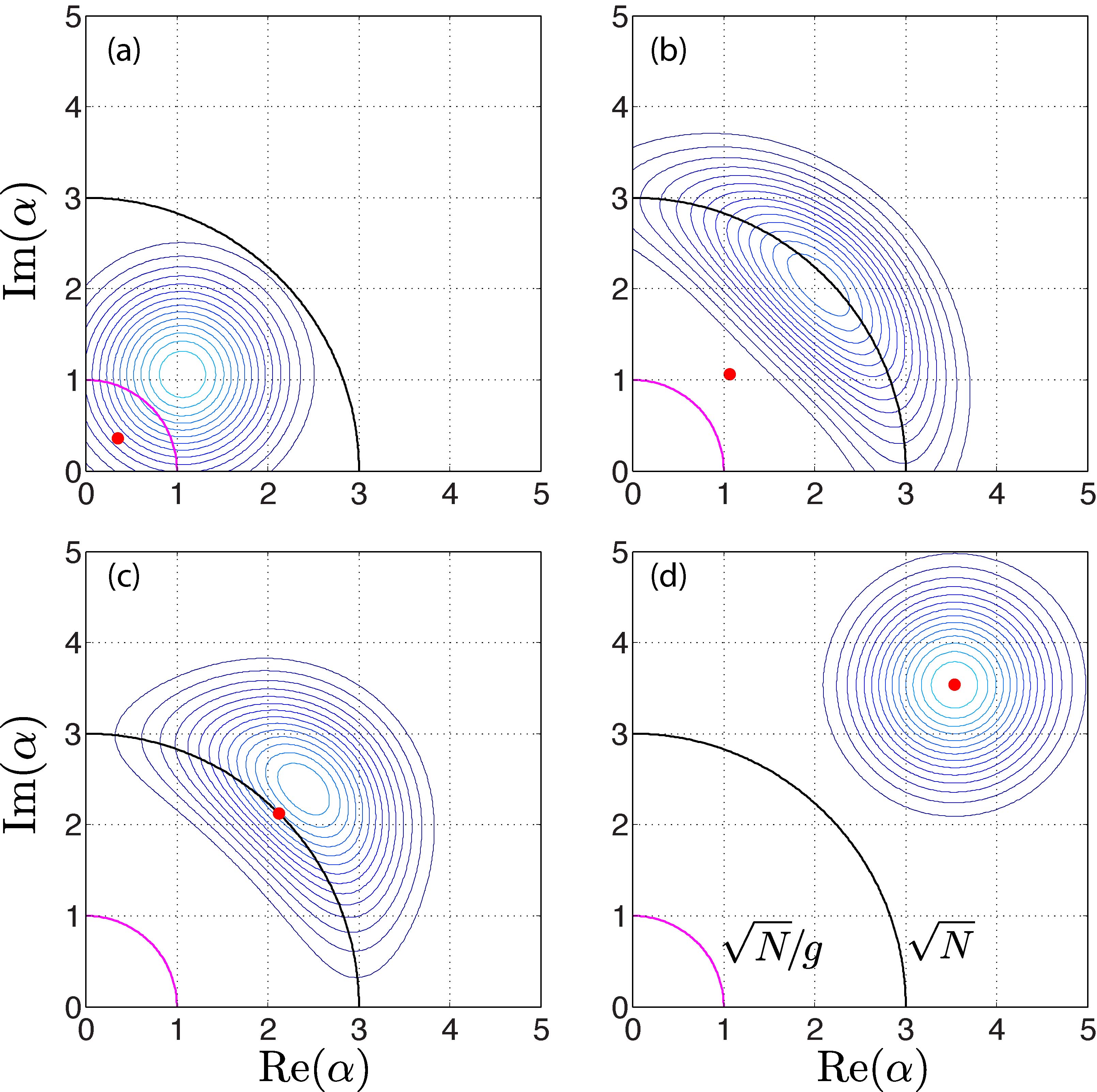}
\caption{(Color online) $Q$ distribution of the output state of the
immaculate linear amplifier given by the extended Kraus operator
$\Upsilon_0$, with $g=3$ and $N=9$, for four amplitudes of input
coherent state: (a)~$\bar\alpha=0.5$; (b)~$\bar\alpha=1.5$;
(c)~$\bar\alpha=3$; (d)~$\bar\alpha=5$.  The (red) dot denotes the
center of the input coherent state.  The transition at input radius
$\sqrt N/g=1$ is marked by a (red) arc, and its image at the output
by the (black) arc at radius $\sqrt N=3$.  Thus (a)~lies within the
high-fidelity region, and the output looks like an amplified coherent
state; (b)~lies beyond the transition, and its output is flattened
along the arc of radius $\sqrt N$.  A second transition occurs near
$\bar\alpha\simeq\sqrt N$, as $\Upsilon_0$ transitions to being the
identity operator.  Thus (c), lying right in the middle of this
second transition, has output that is little amplified and is
flattened along the radial direction, whereas (d), lying well beyond
the second transition, has output that is nearly identical to the
input coherent state.}\label{fig7}
\end{figure}

The key features of the output state of the immaculate amplifier
$\sA_N=\Upsilon_0\odot\Upsilon_0$ are illustrated by the
$Q$-distribution plots in \frf{fig7}.  In Fig.~\ref{fig7}(a), an
input state within the high-fidelity input region is transformed to
an output state that is very close to the target output coherent
state.  In part~(b), however, the input state is beyond the
high-fidelity input region; the output state gets plastered against
the output arc of radius $\sqrt N$, producing a flattening and
distortion along this arc.  This distortion is very much like that
seen in experiments that implement immaculate linear
amplification~\cite{FerBarBla10,FerBlaBar11,ZavFiuBel11,UsuMulWit10,UsugaThesis12}. (It is worth noting that for the unextended Kraus operator
$P_NK_0$, as $\bar \alpha$ increases beyond $\sqrt{N}/g$, the output
state becomes essentially the Fock state $\ket{N}$.)  Parts~(c)
and~(d) illustrate the passage through the second transition at
$\bar\alpha\simeq\sqrt N$, as the action of $\Upsilon_0$ transitions
to being that of the unit operator.

\begin{figure}[htbp]
\includegraphics[width=0.49\textwidth]{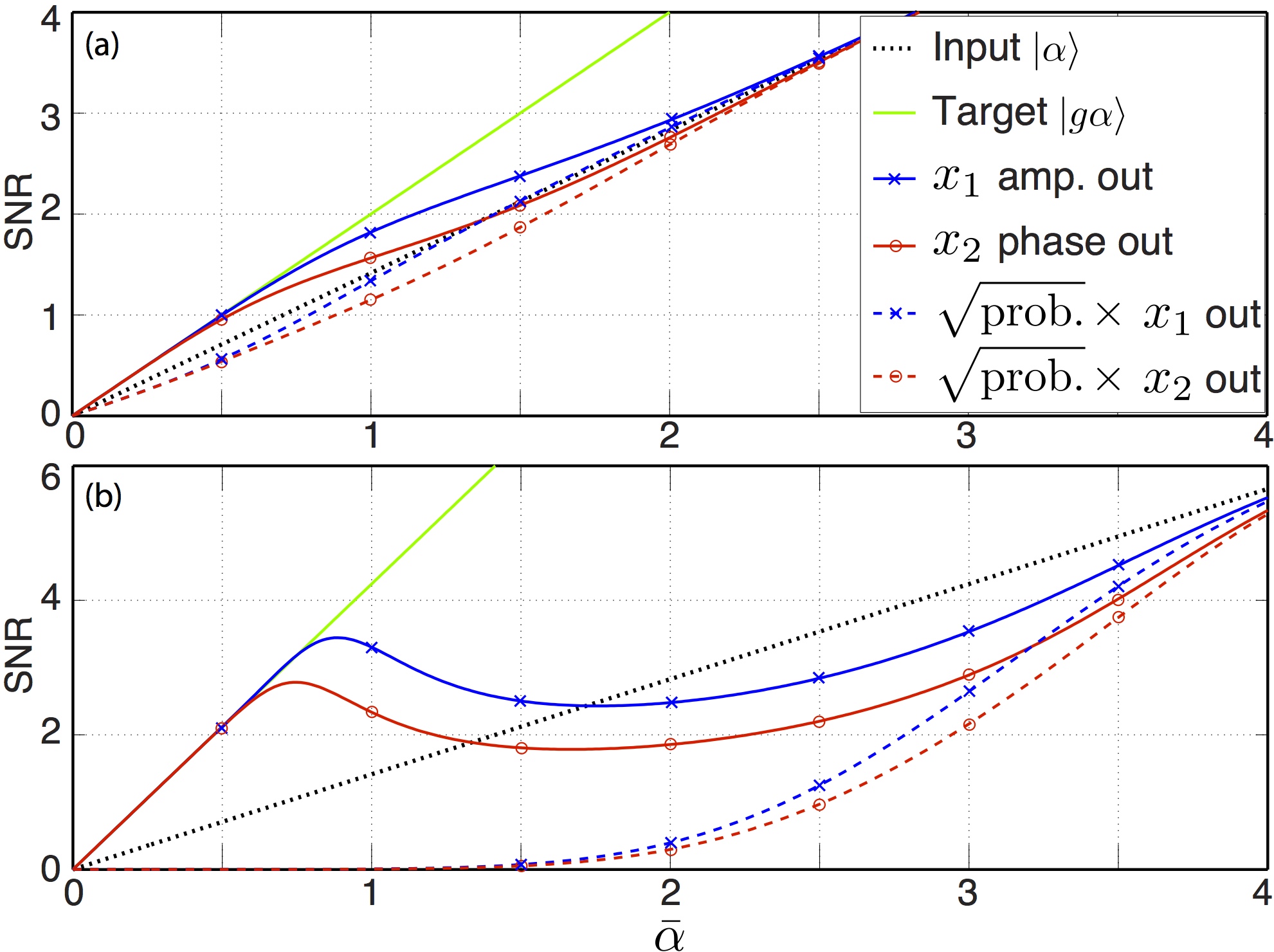}
\caption{(Color online) Antinormally ordered quadrature SNRs as a
function of the input amplitude $\bar \alpha$.  The SNR is defined as
$\SNRone=\expt{x_1}/\delta x_1$ for the amplitude (radial)
quadrature $x_1$ or as $\SNRtwo=\expt{x_1}/\delta x_2$ for
the phase quadrature $x_2$.  Four of the plots are for (i)~the input
state $\ket{\alpha}$ (dotted line), for which $\SNRone=\SNRtwo$ {[this is also the bound given in \erf{eq:SNR_bound}]};
(ii)~the output target state $\ket{g \alpha}$ (solid line), for which
$\SNRone=\SNRtwo$; and~(iii) and~(iv) the output state of the
$\Upsilon_0$ immaculate amplifier (SNR$_1$: solid line with crosses;
SNR$_2$: solid line with circles).  The other two plots give the
amplifier SNRs multiplied by the square root of the working
probability, $\sqrt{p_0(\checkmark|\bar\alpha)}$, as described in the
text: the dashed line with crosses plots
$\sqrt{p_0(\checkmark|\bar\alpha)}\SNRone$, and the dashed line with
circles plots $\sqrt{p_0(\checkmark|\bar\alpha)}\SNRtwo$.  For the
amplifier plots, (a)~has $g=\sqrt{2}$, $N=2$, and (b)~has $g=3$,
$N=9$.} \label{fig8}
\end{figure}

{In Fig.~\ref{fig8} we plot the SNR-based performance measure
defined in Sec.~\ref{subsec:upbounds}, with the key difference that
we have two such SNRs: $\SNRone=\expt{x_1}/\delta
x_1=\sqrt2\bar\alpha/\delta x_1$ for the amplitude (radial)
quadrature $x_1$ and $\SNRtwo=\expt{x_1}/\delta x_2$ for the phase
quadrature $x_2$ ($\langle x_2\rangle=0$).  As in
Sec.~\ref{subsec:upbounds}, the uncertainties in the SNRs are
calculated using antinormal ordering, which applies when one intends
to measure both quadratures~\cite{symmetricSNR}.} Figure~\ref{fig8}
plots the SNR quantities for an input coherent state
$|\alpha\rangle$, the target output state $|g\alpha\rangle$, and the
output of an $\Upsilon_0$ immaculate amplifier.  As discussed in
Sec.~\ref{subsec:upbounds}, the right way to take into account the
success probability of the immaculate amplifier is to multiply the
SNRs by the square root of the working probability; thus
Fig.~\ref{fig8} also shows plots the {root-probability--SNRs}, $\sqrt{p_0(\checkmark|\bar\alpha)}\SNRone$
and $\sqrt{p_0(\checkmark|\bar\alpha)}\SNRtwo$, for the output of the
immaculate amplifier.

Part~(a) of Fig.~\ref{fig8} plots these quantities for a gain typical
of that achieved in experiments.  Part~(b) has a larger gain that
shows the characteristic features of these quantities.  Within the
high-fidelity input region, the output SNRs of the amplifier match
those of the target output state, but they fall away from the target
as $\bar\alpha$ moves out of the high-fidelity region. {The
root-probability--SNRs show that once the success probability is
taken into account, the immaculate amplifier does not do as well as
the input coherent state; it always satisfies the
bound~(\ref{eq:SNR_bound}) and is not even close to the bound in the
high-fidelity region.}

\begin{figure}[htbp]
\includegraphics[width=0.49\textwidth]{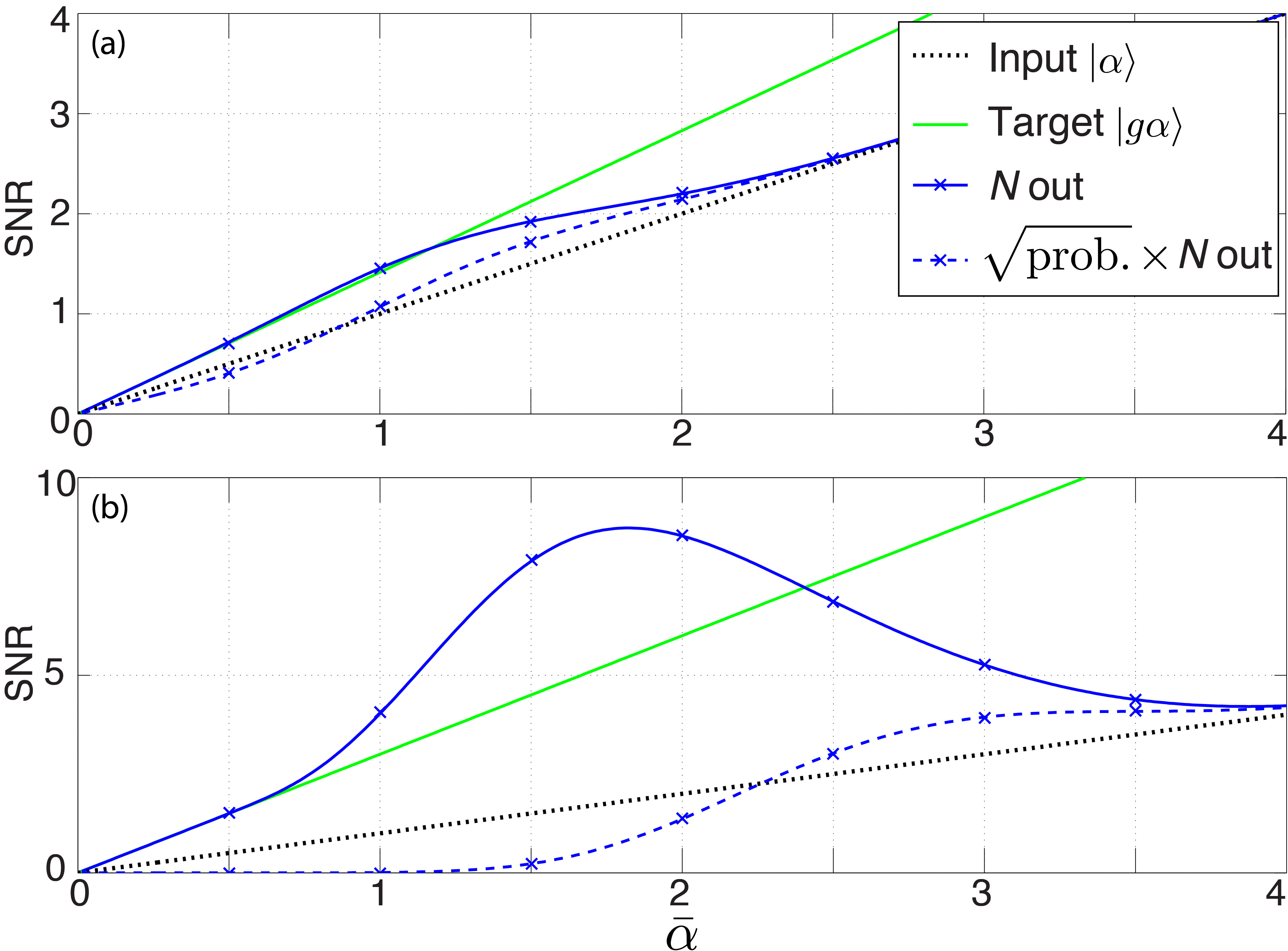}
\caption{(Color online) {Number-based SNR measure as a
function of the input amplitude $\bar \alpha$.  The four plots are
$\SNRnum$ for the input state $\ket{\alpha}$ (dotted line); $\SNRnum$
for the output target state $\ket{g \alpha}$ (solid line); $\SNRnum$
for the output state of the $\Upsilon_0$ immaculate amplifier (solid
line with crosses); and the root-probability--SNR measure
$\sqrt{p_0(\checkmark|\bar\alpha)}\SNRnum$ for the output state of
the $\Upsilon_0$ immaculate amplifier (dashed line with crosses). For
the amplifier plots, (a)~has $g=\sqrt{2}$, $N=2$, and (b)~has $g=3$,
$N=9$.}} \label{fig9}
\end{figure}

{One could use other SNR-based performance measures, an
example being one based on the statistics of number of quanta.  Doing
this can lead to different conclusions.  In \frf{fig9}, we consider a
number-based SNR defined as $\SNRnum =\expt{N}/\Delta N$, where
$N=a\dg a$ is the number operator and $\Delta N$ is the uncertainty
in $N$.  Figure~\ref{fig9} shows that in terms of $\SNRnum$, first,
the output of the immaculate amplifier can do better than the target
output state and, second, the number-based root-probability--SNR,
which includes the square root of the success probability, can exceed
that of the input coherent state.  The first of these improvements
seems to arise from the distortion of the output state as it leaves
the high-fidelity region at $\bar\alpha\simeq\sqrt{N}/g$, which is
$1$ in both plots; this distortion amounts to squeezing in the radial
direction, as is illustrated in \frf{fig7}(b).  The second
improvement is due to the same distortion, but is also aided by the
increase in success probability, displayed in Fig.~\ref{fig6}, for
$\bar\alpha\agt\sqrt N$.  Since these improvements arise from effects
outside the region of high-fidelity immaculate amplification, they
seem to be incidental to the operation of the device as an immaculate
amplifier.}

We conclude this section by re{\"\i}terating that in the
high-fidelity regime, the $k=0$ extended-Kraus-operator immaculate
amplifier has a success probability [see Eq.(\ref{eq:pklimit}) with
$k=0$]
\begin{equation}
p_0(\checkmark|\bar\alpha)=\frac{e^{-\bar\alpha^2}}{g^{2N}}\,.
\end{equation}
This can be regarded as the chief result of this section: {\em within
the high-fidelity region of operation, an approximate
phase-insensitive immaculate linear amplifier has a success
probability that decreases exponentially with the size $N/g^2$ of the
high-fidelity input region, with the base of the exponential being
$g^{2g^2}$.}  This result, for the optimal phase-insensitive
immaculate amplifier, indicates that the very low success
probabilities seen in
experiments~\cite{XiaRalLun10,FerBarBla10,FerBlaBar11}, though they
might be depressed yet further by technical difficulties, are an
unavoidable consequence of trying to perform phase-insensitive
immaculate amplification.

\section{Conclusion}\label{sec:con}

{Immaculate amplification is an attempt to evade the
uncertainty principle by praying that the quantum gods won't be
paying attention all the time.  Unfortunately, our results indicate
that the quantum ones are keenly alert and do not suffer hubris
gladly.  Our chief conclusion is that immaculate amplifiers, if they
operate phase-insensitively, cannot achieve both high fidelity to the
target output state and even reasonably high working probability.
Indeed, in phase-plane regions where a phase-insensitive device
amplifies immaculately with high fidelity, the probability that the
device works is extremely small. The small working probabilities seen
in experiments that implement immaculate amplification are not solely
a consequence of technical imperfections; they are inherent in the
nature of phase-insensitive immaculate amplification.

As penance for trying to outfox quantum mechanics, we suggest several
changes in focus that might reconcile the concept of immaculate
amplification and quantum theory, as well as leading to more positive
results than those reported here.  The first of these is simple:
phase-insensitive immaculate amplification, with its reduction in
noise from input to output, might be a step too far; perhaps a better
sort of device to seek is a probabilistic perfect amplifier, which
would amplify the symmetrically ordered input noise without adding
the noise of a (deterministic) ideal linear amplifier. Working
probabilities for probabilistic perfect amplifiers might be better
than those we have found for immaculate amplifiers.


The second change is to abandon hope for invariance under phase-plane
rotations and of working on more than one circle of input coherent
states, focusing instead on the quite encouraging probabilities we
have found for immaculate amplification of sparse collections of
coherent states on a single input circle.  Nondeterministic devices
have found many uses in quantum information science, a notable
example being the KLM scheme for linear-optical quantum
computing~\cite{Knill01}.  Immaculate amplifiers, like the one
described formally by Eq.~(\ref{eq:Drho}), which are both phase
sensitive and amplitude specific, can work on sparse collections of
coherent states with high success probability; they might find
application in problems such as discrimination of the coherent states
used in phase-shift keying~\cite{NaiYenGuh12,Bec13}.  There are
important questions regarding communications protocols based on such
devices: How robust are they against amplitude and phase noise in the
preparation of the input coherent states?  How badly are rates
impacted by the success probability?  These questions are certainly
worth investigating.

Finally, we suggest a change in the quantum-information-science
approach to analyzing amplifying devices.  The literature on
immaculate amplification has focused on the fidelity of the output
with the immaculate target.  We have stressed that fidelity cannot be
considered as a performance measure alone; the probability-fidelity
product is a better measure of overall performance.  Instead of
attempting to optimize the probability-fidelity product, however, it
might be better to develop performance measures suited to specific
applications.  For metrological applications, the
root-probability--SNR impresses us as an appropriate measure of
performance.  Continuous-variable quantum key distribution is a
communication protocol that might use immaculate amplification and
where key rates are an obvious performance measure.  Some steps have
been taken to optimize key rates in this
context~\cite{FiuCer12,WalRalSym13}, but more work is needed.
To paraphrase Emerson, a foolish fidelity to fidelity
is the hobgoblin of small minds \footnote{The actual quote is \cite{Emerson}: ``A foolish consistency is the hobgoblin of little minds, adored by little statesmen and philosophers and divines.'' };  
that is, each application begs for its own performance measure. 

\begin{acknowledgments}
The authors thank S.~Croke, A.~Denney, F.~Ferreyrol, M.~J.~W. Hall,
S.~Kocsis, A.~P Lund, I.~Marvian, G.~J. Pryde, R.~W. Spekkens, T.~C.
Ralph, and N.~Walk for helpful and enlightening conversations and
e-mail exchanges. This work was supported in part by National Science
Foundation Grant Nos.~PHY-1212445 and PHY-1005540 and by Office of
Naval Research Grant No.~N00014-11-1-0082. JC and CMC also
acknowledge support by the National Science Foundation under Grant
No. NSF PHY-1125915 at the Kavli Institute for Theoretical Physics.
\end{acknowledgments}

\appendix

\section{Linear dependence of coherent states on a circle}
\label{sec:independence_day}

We review the linear dependence of the continuum of coherent states
on a phase-space circle of radius $\bar\alpha$ centered at the
origin.  The reader should also consult the appendix of
Ref.~\cite{DunAnd12}.

A coherent state is represented in the number basis by
\begin{equation}\label{csr}
\ket{\alpha=\bar\alpha e^{i\phi}}
= e^{-\bar\alpha^2/2}
\sum_{n=0}^{\infty}\frac{\bar\alpha^ne^{in\phi}}{\sqrt{n!}}\ket{n}\,.
\end{equation}
The coherent states $|\bar\alpha e^{i\phi}\rangle$, $0\le\phi<2\pi$,
on a circle of radius $\bar\alpha$ are complete, but they are not
linearly independent.

These states are linearly dependent, as we can see from
\begin{equation}\label{ld}
\int_0^{2\pi}\frac{d\phi}{2\pi}\,e^{-in\phi}\ket{\bar\alpha e^{i\phi}}
=
\begin{cases}
\displaystyle{e^{-\bar\alpha^2/2}\frac{\bar\alpha^n}{\sqrt{n!}}|n\rangle}\,,&\mbox{$n\ge0$},\\
0,&\mbox{$n<0$}.
\end{cases}
\end{equation}
The vanishing of the integral for $n<0$ shows the states are not
linearly independent.

That these states are complete follows immediately from expanding any
vector as
\begin{align}
|\psi\rangle=\sum_{n=0}^\infty |n\rangle\langle n|\psi\rangle
=\int\frac{d\phi}{2\pi}\,\chi(\phi)|\bar\alpha e^{i\phi}\rangle\,,
\end{align}
where the function $\chi(\phi)$ has Fourier representation
\begin{align}
\chi(\phi)=\sum_{n=0}^\infty\chi_n e^{-in\phi}\,,
\end{align}
with the positive Fourier coefficients uniquely determined to be
\begin{align}
\chi_n=e^{\bar\alpha^2/2}
\frac{\sqrt{n!}}{\bar\alpha^n}\langle n|\psi\rangle\,,
\quad n>0,
\end{align}
and the negative Fourier coefficients arbitrary.  That the negative
Fourier coefficients can be changed arbitrarily without changing
$|\psi\rangle$ expresses the linear dependence of the coherent states
on a circle.

\vspace{18pt}

\section{Rotationally symmetric quantum operations}\label{sec:phasepreserve}

The superoperator that effects a rotation by $\theta$ in the phase
plane is
\begin{align}
\sR(\theta)=e^{i\theta a^\dagger a}\odot e^{-i\theta a^\dagger a}
=\sum_{n,m}e^{i(n-m)\theta}|n\rangle\langle n|\odot|m\rangle\langle m|\,.
\end{align}
A quantum operation $\sA$ is invariant under rotations if it commutes
with $\sR(\theta)$ for all $\theta$, i.e.,
$\sR(\theta)\circ\sA=\sA\circ\sR(\theta)$.  The symmetry condition
implies that $\sA$ has the form
\begin{align}
\sA=\sum_k\sum_{n,m}A^{(k)}_{nm}|n+k\rangle\langle n|\odot|m\rangle\langle m+k|\,.
\end{align}
That $\sA$ is a quantum operation, i.e., is completely positive,
implies that $A^{(k)}$ is a positive Hermitian matrix and thus can be
diagonalized by a unitary matrix:
\begin{equation}
A^{(k)}_{nm}=\sum_l\lambda^{(k)}_l U_{nl}^{(k)}U^{(k)*}_{ml}\,.
\end{equation}
This brings $\sA$ into the form
\begin{align}
\sA=\sum_{k,l}M^{(k)}_l\odot M^{(k)\dagger}_l\,,
\end{align}
where the operators
\begin{align}
M^{(k)}_l=\sum_n \sqrt{\lambda^{(k)}_l}U^{(k)}_{nl}|n+k\rangle\langle n|
\label{eq:symmKraus}
\end{align}
are orthogonal Kraus operators.  Invariance under rotations manifests
itself as the requirement that these Kraus operators have nonzero
number-basis matrix elements only in one diagonal strip specified by
the integer~$k$.\\ \\

\section{Optimal success probability and fidelity}
\label{sec:optimal}

In this Appendix we show that the success probabilities and
fidelities of Eqs.~(\ref{eq:pkopt2}) and~(\ref{eq:Fkopt}) satisfy the
bounds~(\ref{eq:pkbounds}), (\ref{eq:Fkbounds}),
and~(\ref{eq:F0ineq}).

We first show the inequalities
\begin{align}
\label{eq:pkineq}
p_k(\checkmark|\bar\alpha)&\ge p_{k+1}(\checkmark|\bar\alpha)\,,\\
F_k(\bar\alpha)&\ge F_{k+1}(\bar\alpha)\,;
\label{eq:Fkineq}
\end{align}
from these, we can also conclude that
$p_k(\checkmark|\bar\alpha)F_k(\bar\alpha)\ge
p_{k+1}(\checkmark|\bar\alpha)F_{k+1}(\bar\alpha)$.  This proves that
the best success probability and fidelity are achieved at $k=0$,
i.e., by the Kraus operator $\Upsilon_0$.

The success-probability inequalities~(\ref{eq:pkineq}) follow
straightforwardly from the difference
\begin{widetext}
\begin{align}
Q_k =
\Upsilon\dg_k\Upsilon_k-\Upsilon\dg_{k+1}\Upsilon_{k+1}
=\frac{N!}{(N+k)!}\frac{1}{g^{2N}}
\sum_{n=0}^{N}\frac{(n+k)!}{n!}g^{2n}
\left(1-\frac{n}{N+k+1}\frac{1}{g^2}\right)\op{n+k}{n+k}\ge0
\,.
\end{align}
The manifest positivity of $Q_k$ means that
$\langle\alpha|Q_k|\alpha\rangle\ge0$, which is the
inequality~(\ref{eq:pkineq}).

To show the fidelity inequalities~(\ref{eq:Fkineq}), we begin by
writing Kraus operator~(\ref{eq:Upsilonk}) in the form
\begin{equation}
\Upsilon_k=\sum_{n=0}^\infty
f_k(n)\sqrt{\frac{(n+k)!}{n!}}|n\rangle\langle n+k|\,,
\end{equation}
where
\begin{equation}
f_k(n)=
\begin{cases}
\displaystyle{\sqrt{\frac{N!}{(N+k)!}}\frac{g^n}{g^N}}\,,& n=0,\ldots,N,\\
\displaystyle{\sqrt{\frac{n!}{(n+k)!}}}\,,& n=N+1,N+2,\ldots\,.
\end{cases}
\end{equation}
Notice that $f_k(n)$ does not decrease with $n$ for $n\le N$, reaches
it maximum value at $n=N$, and then is a nonincreasing function of
$n$ for $n\ge N$.

Using $f_k(n)$, we can write
\begin{equation}
\langle g\alpha|\Upsilon_k|\alpha\rangle
=e^{-(g^2-1)\bar\alpha^2/2}\alpha^k\mbox{E}[g^nf_k(n)]\,,
\end{equation}
where $\mbox{E}$ denotes an expectation value with respect to the
Poisson distribution $\mbox{Pr}[\,n\mid\bar\alpha^2\,]=|\langle
n|\alpha\rangle|^2\equiv P_n$.  We also have
\begin{equation}
p_k(\checkmark|\bar\alpha)=\langle\alpha|\Upsilon_k^\dagger\Upsilon_k|\alpha\rangle
=\bar\alpha^{2k}\mbox{E}[f_k^2(n)]\,.
\end{equation}
Thus the fidelity~(\ref{eq:Fkopt}) can be put in the form
\begin{equation}\label{eq::fid_app}
F_k(\bar\alpha)
=e^{-\left(g^2-1\right)|\alpha|^2}
\frac{\bigl(\mbox{E}[g^nf_k(n)]\bigr)^2}{\mbox{E}[f_k^2(n)]}\,.
\end{equation}

For any $k=0,1,\ldots\,$, we define
\begin{align}
h_k(n)\equiv\frac{f_{k+1}(n)}{f_k(n)}
=\begin{cases}
\displaystyle{\frac{1}{\sqrt{N+k+1}}}\,,& n=0,\ldots,N,\\
\displaystyle{\frac{1}{\sqrt{n+k+1}}}\,,& n=N+1,N+2,\ldots\,.
\end{cases}
\end{align}
Notice that $h_k(n)$ is a nonincreasing function of $n$.

The fidelity inequality~(\ref{eq:Fkineq}) equivalent to
\begin{align}
\mbox{LHS}
=\bigl(\mbox{E}&[g^nf_{k+1}(n)]\bigr)^2\mbox{E}[f_k^2(n)]
\le\bigl(\mbox{E}[g^nf_{k}(n)]\bigr)^2\mbox{E}[f_{k+1}(n)^2]
=\mbox{RHS}\,.
\end{align}
Since, by the Schwarz inequality,
\begin{align}
{\rm LHS}\le\mbox{E}[g^nf_{k}(n)]\mbox{E}[g^nf_{k}(n)h_k^2(n)]\mbox{E}[f_k^2(n)]
\equiv\mathcal{I}\,,
\end{align}
we can achieve our objective by showing that
$\mathcal{I}\le\mbox{RHS}$ or, equivalently, that
\begin{align}\label{almostdone}
\mbox{E}[g^nf_{k}(n)h_k^2(n)]\mbox{E}[f_k^2(n)]
\le
\mbox{E}[g^nf_{k}(n)]\mbox{E}[f_{k+1}^2(n)]\,.
\end{align}

Equation~(\ref{almostdone}) can be written as
\begin{equation}\label{eq:idea}
0\ge\sum_{m,n}G(m,n)=\sum_{n=0}^\infty\sum_{m=0}^{n-1}G(m,n)+G(n,m)\,,
\end{equation}
where
\begin{equation}
G(m,n)=P_nP_mf_k(n)f_k(m)h_k^2(m)[g^mf_k(n)-g^nf_k(m)]\,.
\end{equation}
In the final form of Eq.~(\ref{eq:idea}), we use the fact that
$G(n,n)=0$ to exclude the terms along the diagonal from the sum.  Now
what we show is that
\begin{equation}\label{eq:Gsymm}
G(m,n)+G(n,m)=
P_nP_mf_k(n)f_k(m)[g^mf_k(n)-g^nf_k(m)][h_k^2(m)-h_k^2(n)]
\end{equation}
is never positive for $n>m$.   There are three cases to consider.
First, when $m<n\le N$, $h_k(m)=h_k(n)$, so the
quantity~(\ref{eq:Gsymm}) vanishes.  Second, when $m\le N<n$,
$h_k(m)\ge h_k(n)$ and
\begin{equation}
g^mf_k(n)-g^nf_k(m)=g^m\left(f_k(n)-\frac{g^n}{g^N}f_k(N)\right)\le0\,,
\end{equation}
so the quantity~(\ref{eq:Gsymm}) is not positive.  Third, when
$N<m<n$, $h_k(m)\ge h_k(n)$ and $g^mf_k(n)\le g^nf_k(m)$, so the
quantity~(\ref{eq:Gsymm}) is not positive.  This completes the proof
of the inequalities~(\ref{eq:Fkineq}).

Now we establish the bound~(\ref{eq:F0ineq}) by writing the fidelity
$F_0(\bar\alpha)$ of Eq.~(\ref{eq:Fkopt}) as
\begin{equation}
F_0(\bar\alpha)=e^{-g^2\bar\alpha^2}
\frac{\displaystyle{\left(e_N(g^2\bar\alpha^2)
+g^N\sum_{n=N+1}^\infty\frac{g^n\bar\alpha^{2n}}{n!}\right)^2}}
{\displaystyle{e_N(g^2\bar\alpha^2)
+g^{2N}\sum_{n=N+1}^\infty\frac{\bar\alpha^{2n}}{n!}}}
\ge
e^{-g^2\bar\alpha^2}\left(
e_N(g^2\bar\alpha^2)
+g^N\sum_{n=N+1}^\infty\frac{g^n\bar\alpha^{2n}}{n!}\right)
\ge
e^{-g^2\bar\alpha^2}e_N(g^2\bar\alpha^2)
\,,
\end{equation}
\end{widetext}
where the first inequality follows from using $g^{2N}\le g^Ng^n$ in
the denominator.  This establishes the bound~(\ref{eq:F0ineq}).


\begin{thebibliography}{99}
\bibitem{symmetricSNR}
One could use symmetric ordering instead of the antinormal ordering
used in these SNRs.  Symmetric ordering applies when one intends to
measure just one quadrature, instead of both.
\end{thebibliography}
\end{document}